\documentclass{jfm}

\usepackage{graphicx}
\usepackage{natbib}

\ifCUPmtlplainloaded \else
  \checkfont{eurm10}
  \iffontfound
    \IfFileExists{upmath.sty}
      {\typeout{^^JFound AMS Euler Roman fonts on the system,
                   using the 'upmath' package.^^J}%
       \usepackage{upmath}}
      {\typeout{^^JFound AMS Euler Roman fonts on the system, but you
                   dont seem to have the}%
       \typeout{'upmath' package installed. JFM.cls can take advantage
                 of these fonts,^^Jif you use 'upmath' package.^^J}%
      }
  \else
  \fi
\fi

\ifCUPmtlplainloaded \else
  \checkfont{msam10}
  \iffontfound
    \IfFileExists{amssymb.sty}
      {\typeout{^^JFound AMS Symbol fonts on the system, using the
                'amssymb' package.^^J}%
       \usepackage{amssymb}%

      }{}
  \fi
\fi

\ifCUPmtlplainloaded \else
  \IfFileExists{amsbsy.sty}
    {\typeout{^^JFound the 'amsbsy' package on the system, using it.^^J}%
     \usepackage{amsbsy}}
    {\providecommand\boldsymbol[1]{\mbox{\boldmath $##1$}}}
\fi

\newsavebox{\astrutbox}
\sbox{\astrutbox}{\rule[-5pt]{0pt}{20pt}}

\newcommand{\vect}[1]{{ \boldsymbol{#1} }}
\newcommand{\unit}[1]{\hat{\boldsymbol{#1}}}

\newcommand{\normdw} {\| \vect{\Omega} - \unit{k} \|}
\newcommand{\normdwmean} {\| < \vect{\Omega} - \unit{k} > \|}

\newcommand{\pict}[4][5cm]{
\begin{figure}
\centerline{\includegraphics[width=#1]{#2}}
\caption{\footnotesize{#3}}
\label{#4}
\end{figure}}

\title[Precessing rigid ellipsoids]{Precession driven flows in non-axisymmetric ellipsoids}
\author[J. Noir and D. C\'ebron]{J.\ns N\ls O\ls I\ls R\footnote{Email adress for correspondance: jerome.noir@erdw.ethz.ch} \ns \and \ns  D.\ns C\ls \'E\ls B\ls R\ls O\ls N}
\affiliation{Institut f\"ur Geophysik, ETH Z\"urich, Sonneggstrasse 5, Z\"urich, CH-8092, Switzerland.}

\date{2013}
\begin{document}

\maketitle

\begin{abstract}
We study the flow forced by precession in rigid non-axisymmetric ellipsoidal containers. To do so, we revisit the inviscid
and viscous analytical models which have been previously developed for the spheroidal geometry by respectively Poincar\'e [͓Bull.
Astron. \textbf{27}, 321 (1910)] and Busse [J. Fluid Mech. \textbf{33}, 739 (1968)], and, we report the first numerical simulations of flows in such a geometry. In strong contrast with axisymmetric spheroids where the forced flow is systematically stationary in the precessing frame, we show that the forced flow is unsteady and periodic. Comparisons of the numerical simulations with the proposed theoretical model show excellent agreement for both axisymmetric and non-axisymmetric containers. Finally, since the studied configuration corresponds to a tidally locked celestial body such as the Earth's Moon, we use our model to investigate the challenging but planetary relevant limit of very small Ekman numbers and the particular case of our Moon.

\end{abstract}


\section{Introduction}\label{Intr} 

\subsection{General context}

A rotating solid object is said to precess when its rotation axis itself rotates about a secondary axis that is fixed in an inertial frame of reference. The case of a precessing fluid-filled container has been studied for over one century because of its multiple applications. These flows are indeed present in fluid-filled spinning tops \cite[][]{stewartson1959stability}, gyroscopes, \cite[][]{gans1984dynamics} or tanks of spacecrafts \cite[][]{garg1986spacecraft,agrawal1993dynamic}, possibly affecting the spacecraft stability \cite[][]{bao1997stability}. Precession driven flows are also present in planetary fluid layers, such as the liquid core of the Earth \cite[][]{greff1999core} or the Moon \cite[][]{meyer2011precession}, where they possibly participate in the dynamo mechanism generating their magnetic fields \cite[][]{bullard1949magnetic,bondi1953dynamic,malkus1968precession}. These flows may also have an astrophysical relevance, for instance in neutron stars interiors where they can play a role in the observed precession of radio pulsars \cite[][]{glampedakis2009}.

The first theoretical studies considered the case of an inviscid fluid in a spheroidal container \cite[][]{hough1895oscillations,sloudsky1895rotation,Poincare:1910p12351}. Assuming a uniform vorticity, they obtained a solution, the so-called Poincar\'e flow, given by the sum of a solid body rotation and a potential flow. However, the Poincar\'e solution is modified by the apparition of boundary layers, and some strong internal shear layers are also created in the bulk of the
flow \cite[][]{stewartson1963motion,Busse1968}. These viscous effects have
been taken into account as a correction to the inviscid flow in a spheroid, by
considering carefully the Ekman layer and its critical regions
\cite[][]{Busse1968,zhang2010fluid}. Beyond this correction approach, the complete viscous solution, including the fine description of all the flow viscous layers, has recently been obtained in the particular case of a spherical container with weak precession \cite[][]{kida2011steady}.

When the precession forcing is large enough compared to viscous effects, instabilities can occur and destabilize the Poincar\'e flow. First, the Ekman layers can be destabilized \cite[][]{lorenzani2001fluid2} through standard Ekman layer instabilities \cite[][]{lingwood1997absolute,faller1991instability}. In this case, the instability remains localized near the boundaries. Second, the whole Poincar\'e flow can be destabilized, leading to a volume turbulence: this is the
precessional instability \cite[][]{malkus1968precession}. This small-scale intermittent flow confirm the possible relevance of precession for energy dissipation or magnetic field generation, and has thus motivated many studies. Early experimental attempts \cite[][]{vanyo1991geodynamo,vanyo1995experiments} to confirm the theory of \cite{Busse1968} did not give very good results \cite[][]{pais2001precession}. Simulations have thus been performed in spherical containers \cite[][]{tilgner1999magnetohydrodynamic, tilgner2001fluid}, spheres \cite[][]{noir2001numerical}, and finally in spheroidal containers \cite[][]{lorenzani2001fluid,lorenzani2003inertial}, allowing a validation of the theory of \cite{Busse1968}. Experimental confirmation of the theory has then been obtained in spheroids \cite[][]{Noir2003}, a work followed by many experimental studies involving spheres \cite[][]{goto2007turbulence,kida2010super,boisson2012}, spherical containers \cite[][]{triana2012}, but also cylinders \cite[][]{meunier2008rotating,lagrange2008instability,lagrange2011precessional}.

Finally, the dynamo capability of precession driven flows has then been demonstrated in spheres \cite[][]{tilgner2005precession,tilgner2007kinematic}, spheroids \cite[][]{wu2009dynamo} and cylinders \cite[][]{nore2011nonlinear}, allowing the possibility of a precession driven dynamo in the liquid core of the Earth \cite[][]{kerswell1996upper} or the Moon \cite[][]{dwyerNature}.

\subsection{Motivations}

All the previously cited works have considered axisymmetric geometries. However, in natural systems, both the planet rotation and the gravitational tides deform the celestial body into a triaxial ellipsoid, where the
so-called elliptical (or tidal) instability may take place \cite[][]{lacaze2004elliptical,le2010tidal,cebron2010systematic}. Generally speaking, the elliptical instability can be seen as the inherent
local instability of elliptical streamlines \cite[][]
{bayly1986three,waleffe1990three,le2000three}, or as the parametric resonance between two free inertial waves (resp. modes) of the rotating
unbounded (resp. bounded) fluid and an elliptical strain (of azimutal wavenumber $m=2$). Similarly, it has been suggested that the
precession instability comes from the parametric resonance of two
inertial waves with the forcing related to the precession of
azimutal wavenumber $m=1$ in spheroids \cite[][]{kerswell1993instability,wu2009dynamo} and in cylinders \cite[][]
{lagrange2008instability,lagrange2011precessional}. However, the precession instability is also observed in spheres where there is no $m=1$ forcing from the container boundary. It has thus been suggested that the precession instability may be related to another mechanism \cite[][]{lorenzani2001fluid,lorenzani2003inertial}. Clearly, the precise origin of the
precession instability is still under debate, and is beyond the
point of the present work. But since tides and precession are
simultaneously present in natural systems, it seems necessary to
study their reciprocal influence, in presence or not of
instabilities. 

The full problem is thus rather complex, involving non-axisymmetric geometries and three
different rotating frames: the precessing frame, with a period $T_p
\approx 26\, 000$ years for the Earth, the frame of the tidal bulge,
with a period around $T_d \approx 27$ days for the Earth, and the
container or 'mantle' frame, with a period $T_s = 23.9$ hours for the
Earth. Working in a frame where the geometry is at rest is particularly suitable for theoretical and numerical studies, and this is only possible when two of the three problem timescales are equal. The case $T_d=T_p$ where the container is fixed in the precessing frame has already been considered \cite[][]{cebron2010tilt}, with triaxial ellipsoidal (deformable) containers in order to study the interaction between the elliptical instability and the precession. In this work, we rather focus on the case $T_d=T_s$, which corresponds to rigid precessing containers. This model is thus relevant for fluid layers of terrestrial planets or moons locked in a synchronization state (i.e. $T_d=T_s$) such as the liquid core of the Moon.

The paper is organized as follows. In section \ref{sec2}, we define the problem and introduce the theoretical inviscid and viscous models considered in this work. Using non-linear viscous three-dimensional simulations, we then validate successfully in section \ref{sec3} the proposed theoretical viscous model. The results obtained are then discussed (section \ref{sec4}) and applied to the liquid core of the Moon in the conclusion (section \ref{sec5}).

\section{Mathematical description of the problem} \label{sec2}
We consider an incompressible fluid of density $\rho$ and kinematic viscosity $\nu$ enclosed in a triaxial ellipsoid of principal axes $(a,b,c)$. The cavity rotates along its principal axis of length $c$, and precesses along the unit vector $\unit{k}_p$, as illustrated in figure \ref{frames}a. We denote by $\vect{\Omega}_m$ the instantaneous total vector of rotation of the cavity in the frame of inertia, $\vect{\Omega_o}= \Omega_o \unit{k}$ the rotation vector of the cavity in the precessing frame, and $\vect{\Omega_p}=\Omega_p\, \hat{\vect{k}}_p$ the precession vector in the frame of inertia, such that
\begin{eqnarray}
\vect{\Omega}_m=\vect{\Omega_o}+\vect{\Omega_p}
\end{eqnarray} 

\subsection{Frames of reference}
\pict[13cm]{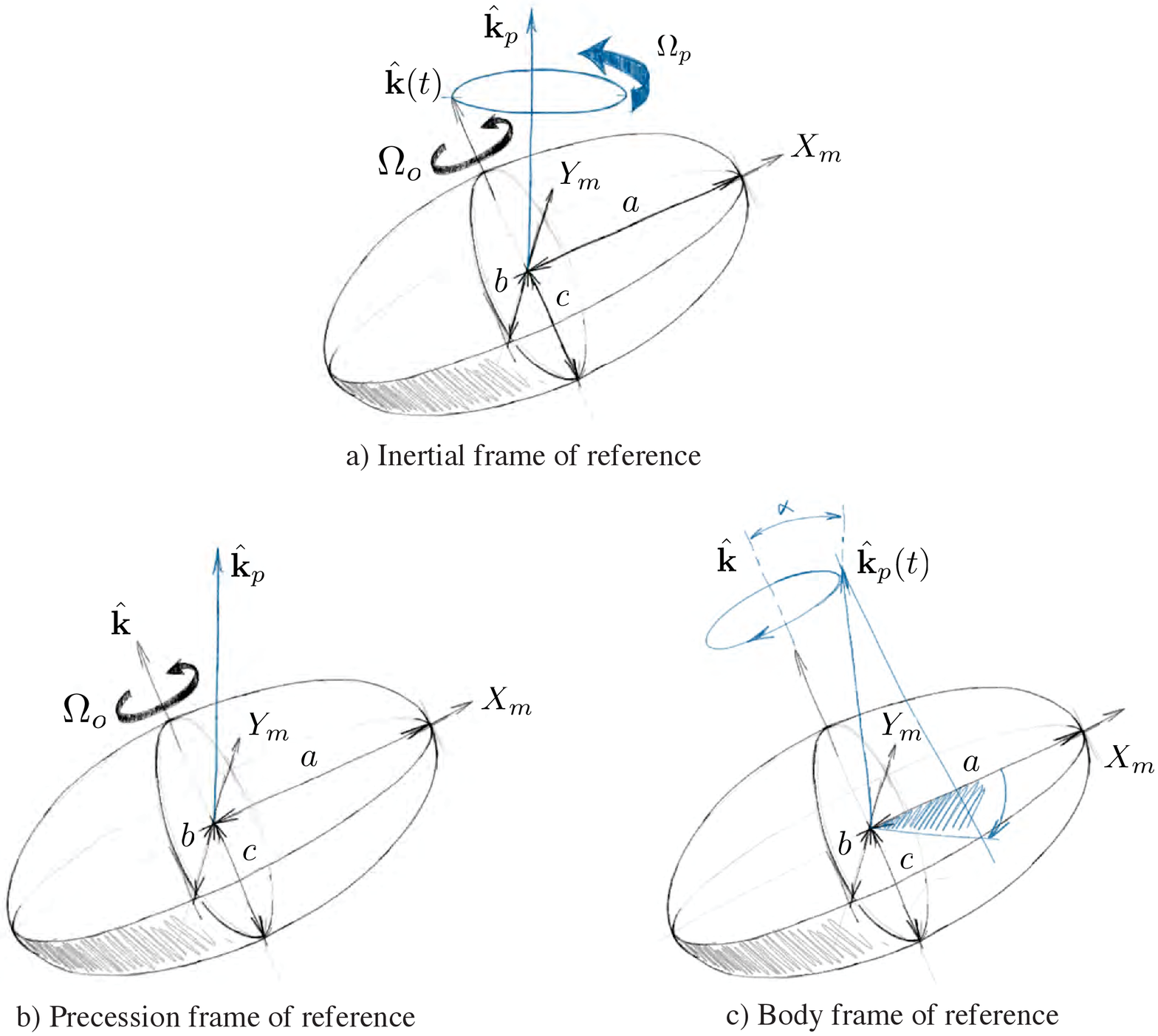}{Schematic representation of the precessing ellipsoidal cavity. a) As seen from the frame of inertia, b) as seen from the frame of precession and c) as seen from the body frame}{frames}
In figure \ref{frames}, we represent the ellipsoidal cavity and the different vectors in the three frames of reference of interest for the present study. In the frame of inertia (fig. \ref{frames}a), the precession vector is fixed, the cavity rotates around the time dependent vector $\hat{\vect{k}}(t)$, which describes a precessional motion around $\hat{\vect{k}}_p$. In the frame of precession (fig. \ref{frames}b), both $\hat{\vect{k}}$ and $\hat{\vect{k}}_p$ are fixed and the cavity rotates around $\hat{\vect{k}}$ (the orientation of the principal axis of the cavity varies in time). In the body frame attached to the cavity (figure \ref{frames}c), the orientation of the principal axes are fixed and the precession vector $\hat{\vect{k}}_p(t)$ exhibits a retrograde motion around $\hat{\vect{k}}$.

\subsection{Coordinate systems}

We define two systems of coordinates (figure \ref{coordinate}): ($\hat{\vect{X}}_m$, $\hat{\vect{Y}}_m$, $\hat{\vect{Z}}_m$), attached to the ellipsoid and oriented along its principal axes $(a,b,c)$, and ($\hat{\vect{X}}_p$, $\hat{\vect{Y}}_p$, $\hat{\vect{Z}}_p$) that is attached to the precessing frame. In the rotating frame attached to the principal axes of the ellipsoid, the unit vectors $\hat{\vect{X}}_p$ and $\hat{\vect{Y}}_p$ rotate in a retrograde direction. We define the time origin such that at $t=0$, $\hat{\vect{X}}_p=\hat{\vect{X}}_m$, $\hat{\vect{Y}}_p=\hat{\vect{Y}}_m$. As shown on figure \ref{coordinate}, $\hat{\vect{Z}}_p=\hat{\vect{Z}}_m$ at all times.

If we consider an arbitrary vector $\vect{A}$ of coordinates ($x_p$, $y_p$, $z_p$) in the coordinates system attached to the precessing frame, its coordinates in the system ($\hat{\vect{X}}_m$, $\hat{\vect{Y}}_m$, $\hat{\vect{Z}}_m$) are given by:
\begin{eqnarray}\label{change_xyz}
x_m&=&x_p \cdot \cos(\Omega_o t)+y_p\cdot\sin(\Omega_o t)\nonumber \\
y_m&=&-x_p\cdot \sin(\Omega_o t)+y_p \cdot \cos(\Omega_o t)\\
z_m&=&z_p \nonumber
\end{eqnarray}
\pict[10cm]{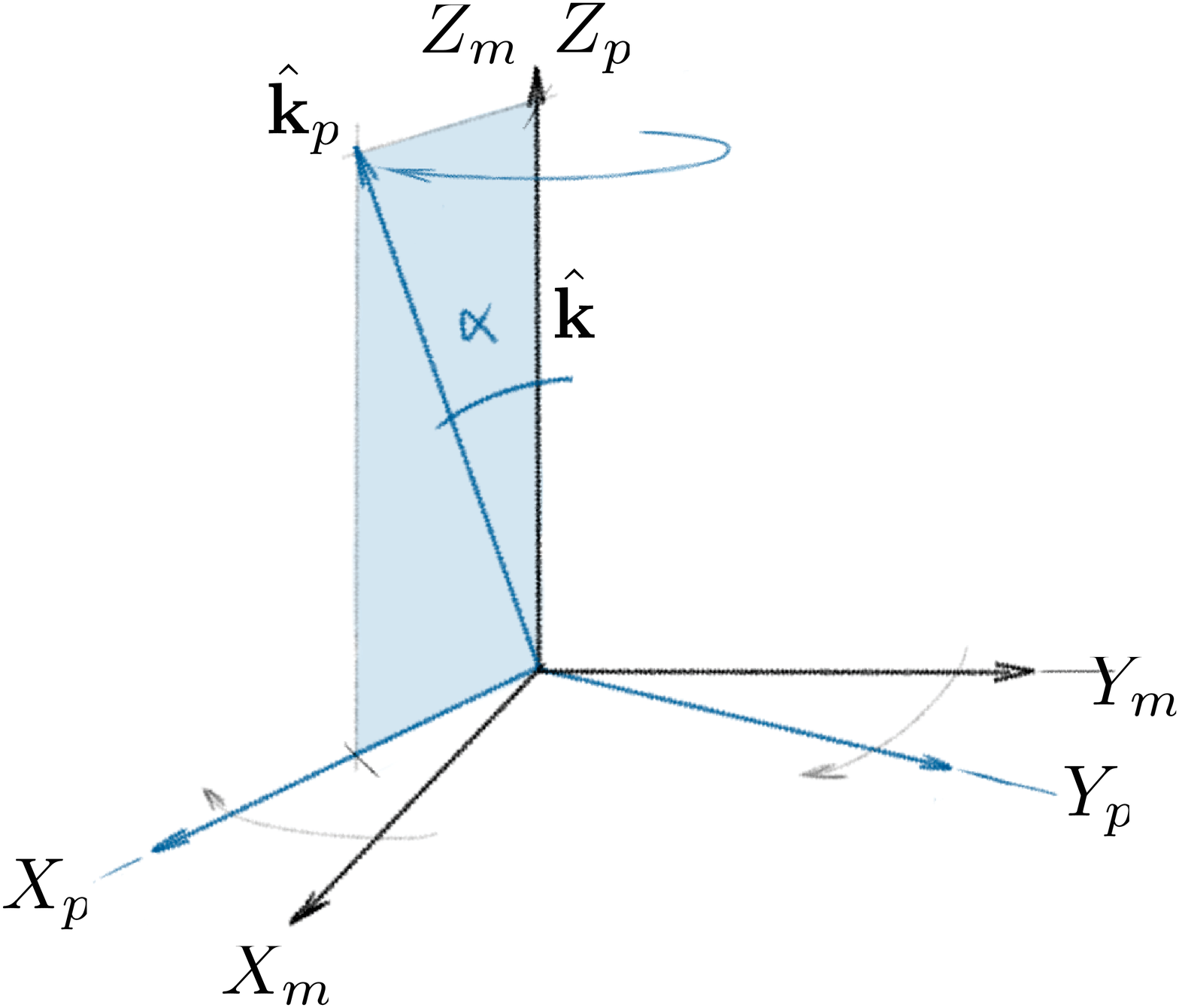}{The coordinate systems: ($\hat{\vect{X}}_m$, $\hat{\vect{Y}}_m$, $\hat{\vect{Z}}_m$) is attached to the body and oriented along the principal axis (a,b,c) of the ellipsoid. In contrast, ($\hat{\vect{X}}_p$, $\hat{\vect{Y}}_p$, $\hat{\vect{Z}}_p$) is attached to the precessing frame.}{coordinate}

In the frame of reference attached to the moving body, the equation of the triaxial ellipsoid boundary is given by:
\begin{eqnarray}
\frac{x^2}{a^2}+\frac{y^2}{b^2}+\frac{z^2}{c^2}=1
\end{eqnarray}
In the present study, we will mostly consider two type of geometries, an axisymmetric spheroid ($a=b\neq c$), to which we refer as a spheroid, and a biaxial ellipsoid ($a\neq b=c$) to which we refer as the non axisymmetric ellipsoid. The true ellipsoidal geometry ($a\neq b \neq c$) will be referred to as the triaxial ellipsoid but will only be considered to derive the general inviscid equations, the fundamental dynamics due to a non axisymmetric equator being already captured when ($a\neq b=c$). The reduced tensor of inertia expressed in the coordinate system attached to the principal axes of the cavity reads
\begin{eqnarray}\label{tensoreinertia}
I=
 \frac{4\pi}{15}   \left [
   \begin{array}{ccc}
      b^2+c^2 & 0& 0 \\
      0& a^2+c^2 & 0 \\
      0& 0& b^2+a^2 \\
   \end{array}
   \right ].
\end{eqnarray}

\subsection{Fluid equations of motion}
Without further assumptions, the fluid motion inside the precessing ellipsoid is governed by the non-linear Navier-Stokes equation. Using $\Omega_o^{-1}$ as a time scale and $R=(abc)^{1/3}$ as a length scale, any velocity field $\vect{u}$ within the precessing ellipsoid is governed by the following equations, expressed in the body frame:
\begin{eqnarray}
\frac{\partial \vect{u}}{\partial t} +2(\unit{k}+Po\hat{\vect{k}}_p)\times\vect{u}+\vect{u}\cdot\nabla \vect{u} &=& -\nabla p - Po(\hat{\vect{k}}_p\times \hat{\vect{k}})\times\vect{r}+E\, \Delta \vect{u}, \label{NS} \\
\nabla \cdot \vect{u} &=& 0, \label{NS2} 
\end{eqnarray}
where $p$ is the reduced pressure which takes the centrifugal force into account, $Po=\Omega_p / \Omega_o$ is the so-called Poincar\'e number, and $E=\nu/(\Omega_o R^2)$ is the Ekman number which represents the relative amplitude of the viscous and Coriolis forces.

If the fluid is viscous, the boundary condition is
\begin{eqnarray}
\vect{u}=\vect{0}, \label{BCNS} 
\end{eqnarray}
which reduces to
\begin{eqnarray}
\vect{u} \cdot \vect{n} = 0 \label{BCNSi} 
\end{eqnarray}
for an inviscid fluid ($E=0$), with $\vect{n}$ the unit vector normal to the boundary surface.

Finally, we introduce the Rossby number that combines the rate of precession $Po$ and the angle of precession $\alpha$, it is a measure of the amplitude of the forcing:
\begin{equation}\label{Rossby}
Ro=Po || \unit{k}_p\times \unit{k} ||=Po\sin\alpha
\end{equation}
 
In former studies, the angle of precession $\alpha$ was fixed and the Poincar\'e number was varied, so does the Rossby number. In this study, we fix the Rossby number, $Ro=10^{-2}$, to ensure that the flow remains stable in our simulations, even at the largest values of $Po$. Consequently, the angle of precession varies as $\sin\alpha=Ro/Po$ as we scan in $Po$. It follows that there is a forbidden band $-10^{-2}<Po<10^{-2}$ for which no $\alpha$ can satisfy (\ref{Rossby}). The same study could be carried out at fixed $\alpha$ by varying the Poincar� number the same conclusions would apply as long as no instability develop in the system. This means that there would be a forbidden zone depending on the critical values of Po, which we do not know. Note finally that to access the small Po region of the parameter space, one could simply reduce accordingly Ro.

\subsection{Inviscid flows of uniform vorticity in triaxial precessing ellipsoids}\label{UVE}

Following the precursory work of \cite{hough1895oscillations}, \cite{sloudsky1895rotation} and \cite{Poincare:1910p12351} we assume the fluid to be inviscid and search for a solution of the velocity that is linear in the spatial coordinates $(x,y,z)$, i.e. a particular solution $\vect{U}$ of uniform vorticity 
\begin{eqnarray}\label{univort}
\vect{U}=\vect{\omega}\times \vect{r}+\nabla \psi,
\end{eqnarray}
where $\vect{\omega}$ is the mean rotation component of the flow and $\nabla \psi$ is the gradient flow needed to satisfy the non-penetration boundary condition. It is straightforward to show that such a solution does not generate any viscous force in the interior, which is consistent with our assumption. 

Taking the curl of the Navier-Stokes equation (\ref{NS}) in the body frame, we obtain the vorticity equation for the particular flow (\ref{univort}):
\begin{eqnarray}\label{vorticity_eq}
\frac{\partial \vect{\omega}}{\partial t}=\left(  \vect{\omega}+Po\, \hat{\vect{k}}_p(t)+\hat{\vect{k}}\right)\cdot \nabla \vect{U}-Po\, \hat{\vect{k}}_p(t)\times\hat{\vect{k}},
\end{eqnarray}

An important step to establish the general equation for the mean vorticity is to express $\vect{U}$, or equivalently $\psi$, as a function of ($\omega,a,b,c,x,y,z$). This can be done by imposing the non-penetration condition on the velocity together with the condition of incompressibility, but this is rather lengthy. Instead, we propose to follow an approach similar to the original work of \citet{Poincare:1910p12351}. 

First, we introduce a geometrical transformation that applies in the body frame where the ellipsoid is fixed and that transforms the triaxial cavity $(a,b,c)$ into a sphere with a unit radius (fig. \ref{sphere_trans}). Using a prime to denote quantities in the spherical domain and no prime for quantities in the true ellipsoid, we have:
\begin{eqnarray}\label{tranformX}
x\rightarrow x'=x/a,\,\,y\rightarrow y'=y/b,\,\,z\rightarrow z'=z/c. 
\end{eqnarray}
The velocity is transformed following the same rules:
\begin{eqnarray}\label{tranformV}
u_x\rightarrow u_x'=u_x/a,\,\,u_y\rightarrow u_y'=u_y/b,\,\,u_z\rightarrow u_z'=u_z/c. 
\end{eqnarray}
It is easy to show that the fluid in the spherical domain remains incompressible, of uniform vorticity, and does not penetrate the boundary.
Note however, that it  does not satisfy the "no slip" nor the "stress
free" boundary conditions. Therefore it can only be a solution of the inviscid Euler equation \citep{Tilgner:1998p13224}. We now make use of this transformation and its reciprocal to easily obtain the analytical expression of the uniform vorticity flow in the body frame of the true ellipsoid. 
%
\pict[10cm]{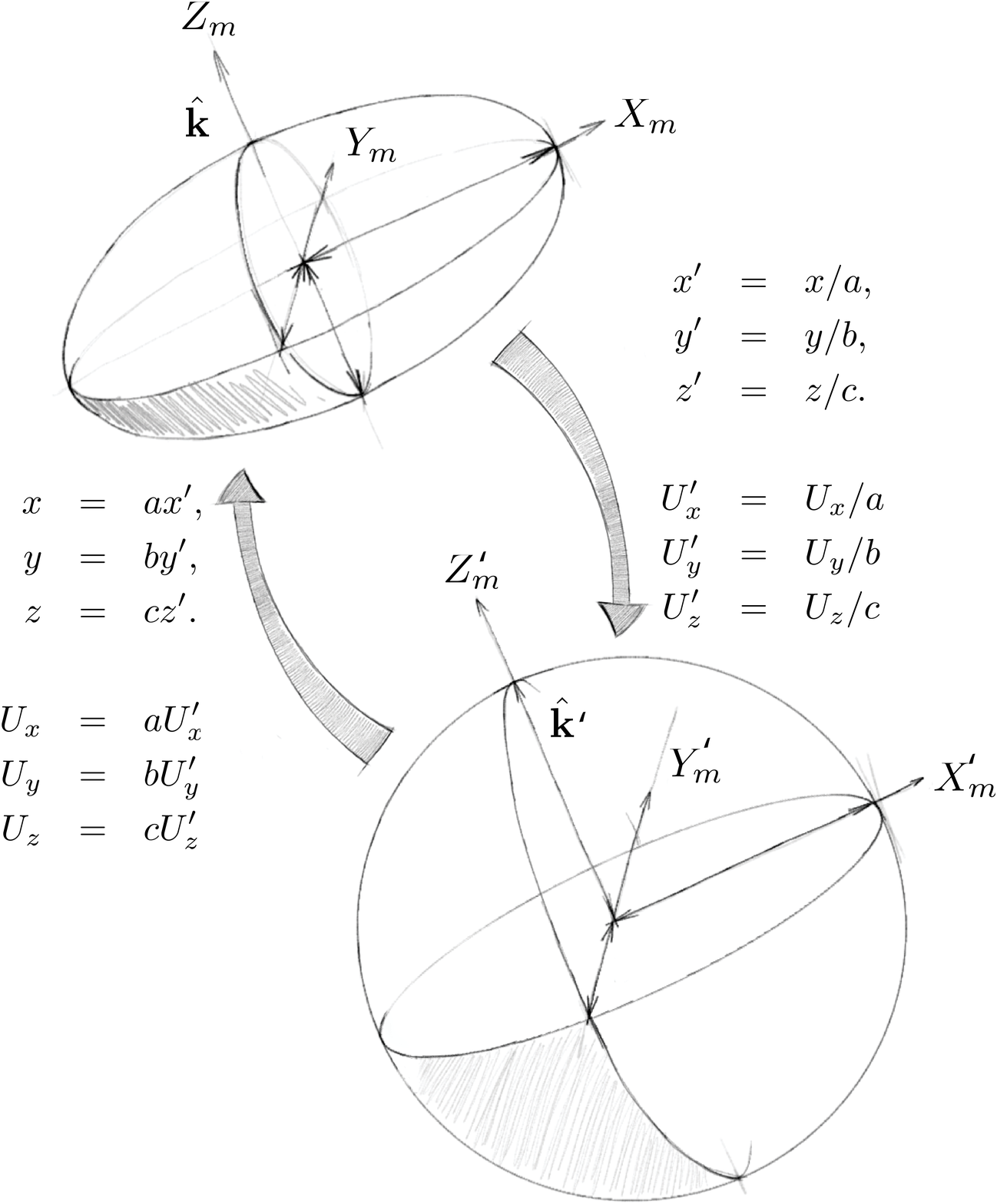}{Schematic representation of the geometrical stretch that transforms a triaxial ellipsoid into a sphere of radius unity.}{sphere_trans}

In the spherical domain, a flow of uniform vorticity simply takes the form of a solid body rotation:
\begin{eqnarray}\label{Uprime}
\vect{U}'=\vect{\omega}'\times \vect{r}'=(\omega'_y z' - \omega'_z y',\, \omega'_z x' - \omega'_x z',\, \omega'_x y' - \omega'_y x').
\end{eqnarray}
Substituting (\ref{tranformX}-\ref{tranformV}) into (\ref{Uprime}) leads to
\begin{eqnarray}\label{velocityU}
\vect{U}=
\left(
\omega'_y \frac{z}{c} - \omega'_z \frac{y}{b},\,\omega'_z \frac{x}{a} - \omega'_x \frac{z}{c},\,\omega'_x \frac{y}{b} - \omega'_y \frac{x}{a}
\right).
\end{eqnarray}
Since $\vect{\omega}'$ is a uniform vector field, the mean vorticity in the ellipsoid is
\begin{eqnarray}\label{vorticityU}
\vect{\nabla}\times\vect{U}=\left(
\omega'_x \left(\frac{c}{b} +\frac{b}{c}\right),\, \omega'_y \left(\frac{a}{c} +\frac{c}{a}\right),\,\omega'_z \left(\frac{b}{a} +\frac{a}{b}\right)
\right)=2\vect{\omega}.
\end{eqnarray}
From (\ref{velocityU}) and (\ref{vorticityU}), we finally obtain the analytical form of uniform vorticity inviscid flows in triaxial ellipsoids:
\begin{eqnarray}\label{velocityabc}
U_x & =& \omega_{y}\,  \frac{2a^2}{a^2+c^2}\, z - \omega_{z}\, \frac{2a^2}{a^2+b^2}\, y,\nonumber\\
U_y & =& \omega_{z}\, \frac{2b^2}{a^2+b^2}\, x - \omega_{x}\, \frac{2b^2}{c^2+b^2}\, z,\\
U_z & =& \omega_{x}\, \frac{2c^2}{b^2+c^2}\, y - \omega_{y}\, \frac{2c^2}{a^2+c^2}\, x.\nonumber
\end{eqnarray}
Identifying the terms within (\ref{univort}), we obtain the expression for the potential field $\psi$:
\begin{eqnarray}
\psi=\omega_{x}\,  \frac{c^2-b^2}{c^2+b^2}\,  yz + \omega_{y}\,  \frac{a^2-c^2}{a^2+c^2}\,  xz + \omega_{z}\,  \frac{b^2-a^2}{b^2+a^2}\,  xy.
\end{eqnarray}
Anticipating the rest of the paper we introduce, $\vect{\Omega}$, the space averaged rotation vector of the fluid in the precessing frame:
\begin{eqnarray}\label{OmegaPrec}
\vect{\Omega}=\vect{\omega}+\hat{\vect{k}}.
\end{eqnarray}
Using the coordinate system attached to the principal axes of the ellipsoid, we obtain from (\ref{change_xyz})
\begin{eqnarray}\label{change_vort}
\omega_{x}&=&\Omega_{x}\cos(t)+\Omega_{y}\sin(t),\nonumber\\
\omega_{y}&=&-\Omega_{x}\sin(t)+\Omega_{y}\cos(t),\\
\omega_{z}&=&\Omega_{z}-1.\nonumber
\end{eqnarray}

Substituting the analytical expression of the velocity (\ref{velocityabc}) in the vorticity equation (\ref{vorticity_eq}), we obtain the general form of the equations that govern the inviscid solution of uniform vorticity in the body frame for a precessing triaxial ellipsoid:
\begin{eqnarray}\label{meanvorteq}
\frac{\partial \omega_{x}}{\partial t} = 2a^2\, \left[ \frac{1}{a^2+c^2}-\frac{1}{a^2+b^2}\right]\omega_{z} \omega_{y} + P_x\sin(t)\frac{2a^2}{a^2+b^2}\omega_{z}\cr
+\left( P_z+1 \right)\frac{2a^2}{a^2+c^2}\omega_{y}+P_x\sin(t), \label{meanvorteq1} \\
\frac{\partial \omega_{y}}{\partial t} = 2b^2\, \left[ \frac{1}{a^2+b^2}-\frac{1}{b^2+c^2}\right]\omega_{x} \omega_{z} + P_x\cos(t)\frac{2b^2}{a^2+b^2}\omega_{z} \cr
-\left( P_z+1 \right)\frac{2b^2}{b^2+c^2}\omega_{x}+P_x\cos(t),\label{meanvorteq2} \\
\frac{\partial \omega_{z}}{\partial t} = 2c^2\, \left[ \frac{1}{b^2+c^2}-\frac{1}{a^2+c^2}\right]\omega_{x} \omega_{y} - P_x\cos(t)\frac{2c^2}{a^2+c^2}\omega_{y} \cr
- P_x\sin(t)\frac{2c^2}{b^2+c^2}\omega_{x}, \label{meanvorteq3}
\end{eqnarray}
with $P_x=Po \sin\alpha=Ro$ and $P_z=Po \cos\alpha=Po\sqrt{Ro^2-Po^2}$. These equations are valid for any values of $(a,b,c)$, $Po$ and $\alpha$. In a spheroidal cavity there exist an infinite number of stationary solutions for the system (\ref{meanvorteq1}-\ref{meanvorteq3}) given by:
\begin{equation}
\vect{\omega}+\unit{k}=\xi \unit{k}_p,
\end{equation}
where $\xi$ can be any real number. Among this class of inviscid solutions, only the solution $\xi=-Po$ remains stationary when $a \neq b$, for all $c$.

\subsection{Reintroducing the viscosity}
Without viscous damping, the inviscid solutions depend on the initial conditions and are somewhat of limited interest. Assuming the Ekman number to be small, we reintroduce the viscosity through the viscous torque due to the friction in the Ekman boundary layer.

In appendix \ref{appendixA}, we extend the approach of \cite{Noir2003} for a spheroid to the case of finite ellipticity. Without any loss of generality, the viscous equations (\ref{meanvorteq1}-\ref{meanvorteq3}) can be written as 
\begin{eqnarray}\label{viscmeanvorteq}
\frac{\partial \omega_{x}}{\partial t}=\left[ \frac{2a^2}{a^2+c^2}-\frac{2a^2}{a^2+b^2}\right]\omega_{z} \omega_{y} + P_x\sin(t)\frac{2a^2}{a^2+b^2}\omega_{z} \cr
+\left( P_z+1 \right)\frac{2a^2}{a^2+c^2}\omega_{y}+P_x\sin(t)+ \left. \mathcal{L} \vect{\Gamma}_{\nu}\right|_x, \label{viscmeanvorteq1} \\
\frac{\partial \omega_{y}}{\partial t}=\left[ \frac{2b^2}{a^2+b^2}-\frac{2b^2}{b^2+c^2}\right]\omega_{x} \omega_{z} + P_x\cos(t)\frac{2b^2}{a^2+b^2}\omega_{z} \cr
-\left( P_z+1 \right)\frac{2b^2}{b^2+c^2}\omega_{x}+P_x\cos(t)+ \left. \mathcal{L} \vect{\Gamma}_{\nu}\right|_y,\label{viscmeanvorteq2} \\
\frac{\partial \omega_{z}}{\partial t}=\left[ \frac{2c^2}{b^2+c^2}-\frac{2c^2}{a^2+c^2}\right]\omega_{x} \omega_{y} - P_x\cos(t)\frac{2c^2}{a^2+c^2}\omega_{y} \cr
- P_x\sin(t)\frac{2c^2}{b^2+c^2}\omega_{x}+ \left. \mathcal{L} \vect{\Gamma}_{\nu}\right|_z.\label{viscmeanvorteq3}
\end{eqnarray}
Using the linear asymptotic of spin-up and of the spin-over mode, we derive an analytical expression of the viscous term in the limit of small Ekman number. 
 
\begin{eqnarray}\label{generalizedvisc}
\mathcal{L} \vect{\Gamma}_{\nu}= \sqrt{E\Omega} \left[ 
\frac{\lambda^r_{so}}{\Omega^2}\left( 
 \begin{array}{c}
      \Omega_x\Omega_z \\
      \Omega_y\Omega_z\\
      \Omega_z^2-\Omega^2\\
   \end{array}
   \right)+
 \frac{\lambda^i_{so}}{\Omega}\left(   
  \begin{array}{c}
     \Omega_y\\
     -\Omega_x\\
      0\\
   \end{array}
   \right)
   +
   \lambda_{sup} \frac{\Omega^2-\Omega_z}{\Omega^2} \left( 
 \begin{array}{c}
      \Omega_x\\
     \Omega_y\\
    \Omega_z\\
   \end{array}
   \right)
\right],
\end{eqnarray}

where $\lambda_{so}^{r}$ and $\lambda_{so}^{i}$ represents the decay rate and viscous correction to the eigenfrequency of the spin-over mode, respectively. In the context of spheroids of finite ellipticity, we use the asymptotic values derived by \citet{Zhang2004}. $\lambda_{sup}$ is an integrated value of the spin-up decay rate and is derived from the asymptotic theory of \citet{Greenspan1968}. We refer to this form of the viscous term as the generalized model in the rest of the paper

In the case of a non-axisymmetric container, no analytical solution for the inertial modes exists. In the lack of a proper theory for the viscous damping of inertial modes in a non axisymmetric container, we adopt the following reduced form for the viscous torque
\begin{eqnarray}\label{addhocvisc}
\mathcal{L} \vect{\Gamma}_{\nu}= \lambda \sqrt{E} \left( 
 \begin{array}{c}
      \Omega_x\\
      \Omega_y\\
      \Omega_z-1\\
   \end{array}
\right).
\end{eqnarray}

In appendix \ref{compare_model_axi}, we show that, for an axisymmetric container, the viscous term (\ref{generalizedvisc}) is well approximated by the reduced form (\ref{addhocvisc}) in the range of parameters considered in this study. Hence, $\lambda$ can be interpreted as an approximation of the decay rate $\lambda^r_{so}$ of the spin-over mode when the contribution from the terms proportional to $\lambda^i_{so}$ and $\lambda_{sup}$ are negligible. In the absence of model of the spin-over mode in a non-axisymmetric ellipsoid, $\lambda$ in our model remains an adjustable parameter and is determined so as to best fit the numerical results in each geometry. Herein, we refer to the viscous set of equation using (\ref{addhocvisc}) as the reduced model.
 
Anticipating the rest of the paper, we introduce the reduced viscous equations in the frame of precession for the particular class of ellipsoid ($a \neq b=c$). Substituting (\ref{change_xyz}) into the inviscid set of equations (\ref{viscmeanvorteq1}-\ref{viscmeanvorteq3}) with $b=c$, we obtain:
\begin{eqnarray}
\frac{\partial \Omega_x}{\partial t}=P_z\Omega_y&+&(1-\chi)\left[ \cos(2t) \left(\frac{P_z+2}{2}\Omega_y -\frac{1}{2}\Omega_y\Omega_z\right)\right.\nonumber \\
&+&\sin(2t)\left( -\frac{P_z+2}{2}\Omega_x +P_x\Omega_z + \frac{1}{2}\Omega_x\Omega_z-P_x\right)\nonumber \\
&+&\left.\frac{P_z}{2}\Omega_y+\frac{1}{2}\Omega_y\Omega_z \right]+\lambda \sqrt{E}\Omega_x\label{eq_Naxi_P_x}, \\
\frac{\partial \Omega_y}{\partial t}=P_x\Omega_z-P_z\Omega_x&+&(1-\chi)\left[ \cos(2t) \left(\frac{P_z+2}{2}\Omega_x -\frac{1}{2}\Omega_x\Omega_z-P_x\Omega_z+P_x\right)\right.\nonumber \\
&+&\sin(2t)\left( \frac{P_z+2}{2}\Omega_y - \frac{1}{2}\Omega_y\Omega_z\right)\nonumber \\
&-&\left.\frac{P_z}{2}\Omega_x-\frac{1}{2}\Omega_x\Omega_z \right]+\lambda \sqrt{E}\Omega_y\label{eq_Naxi_P_y}, \\
\frac{\partial \Omega_z}{\partial t}=-P_x\Omega_y&+&(1-\chi)\left[ \cos(2t) \left(\frac{P_x}{2}\Omega_y+\Omega_x\Omega_y\right)\right.\nonumber \\
&+&\sin(2t)\left( -\frac{P_x}{2}\Omega_x+\frac{1}{2}\left( \Omega_y^2-\Omega_x^2\right)\right)+\left.\frac{P_x}{2}\Omega_y \right]\nonumber \\
&+&\lambda \sqrt{E}(\Omega_z-1)\label{eq_Naxi_P_z},
\end{eqnarray}
with the ratio $\chi$ of the two equatorial moments of inertia:
\begin{eqnarray}
\chi=\frac{b^2+c^2}{a^2+b^2}=\frac{2b^2}{a^2+b^2}.
\end{eqnarray}

\section{Comparison of the theoretical models with numerical simulations} \label{sec3}
To allow for an easy comparison with former studies we will focus our diagnostic on two quantities, the rotation vector of the fluid viewed from the frame of precession, $(\Omega_x,\Omega_y,\Omega_z)$, and the amplitude of the differential angular velocity between the fluid and the surrounding container, $\normdw$
\subsection{Methods} \label{num}

The system of ordinary differential equations describing the time evolution of the uniform vorticity components of the flow, i.e. equations (\ref{viscmeanvorteq1}-\ref{viscmeanvorteq3}), is solved using the Dormand-Prince method, the standard version of the Runge-Kutta algorithm implemented in MATLAB. We have checked that the time evolution is not modified by the use of other time-stepping solvers.

The system of partial differential equations of the initial viscous problem, i.e the equations of motion (\ref{NS}-\ref{NS2}) completed by the boundary condition (\ref{BCNS}) at the ellipsoid surface, is solved using a finite element method implemented in the commercial code COMSOL Multiphysics\textsuperscript{\circledR}. The mesh element type used for the fluid variables is the standard Lagrange element
$P1-P2$, which is linear for the pressure field and quadratic for
the velocity field. For time-stepping, we use the Implicit Differential-Algebraic solver (IDA solver), based on variable-coefficient Backward Differencing Formulas or BDF \cite[see][for details on the IDA solver]{hindmarsh2005sundials}. The integration method in IDA is variable-order, the order ranging between 1 and 5. At each time step the system is solved with the sparse direct linear solver PARDISO (www.pardiso-project.org). All computations have been performed on a single workstation.

Since we are concerned with the effect of topography in our system, we have chosen to fix the Ekman number, $E=10^{-3}$, which allow us to use meshes with typically $30\, 000$ degrees of freedom. Convergence tests in a spherical geometry have been performed to ensure that our simulations with this resolution capture correctly the viscous effects due the Ekman boundary layer. Figure \ref{results_sphere} represents the norm of differential rotation between the fluid and the container in a spherical geometry, $\normdw$. The red symbols represents the numerical simulations, the red dashed line represents the asymptotic solution of \citet{Busse1968}, the dashed blue line represents the generalized model (\ref{eq_axi_P_x}- \ref{eq_axi_P_z}) and the green dashed line represents the reduced model. The best fit leads to $\lambda=-2.62$. We observe a quantitative agreement between all the models and the numerical simulations. A close look at the critical Po shows that the reduced model predicts a resonance at zero while the generalised model and Busse's theory predicts a resonance at $Po\sim -0.01$. This difference is consistent with the fact that the reduced model does not account for the viscous correction of the eigenfrequency of the Poincar� mode, which at these parameters is of order $0.01$.

\pict[10cm]{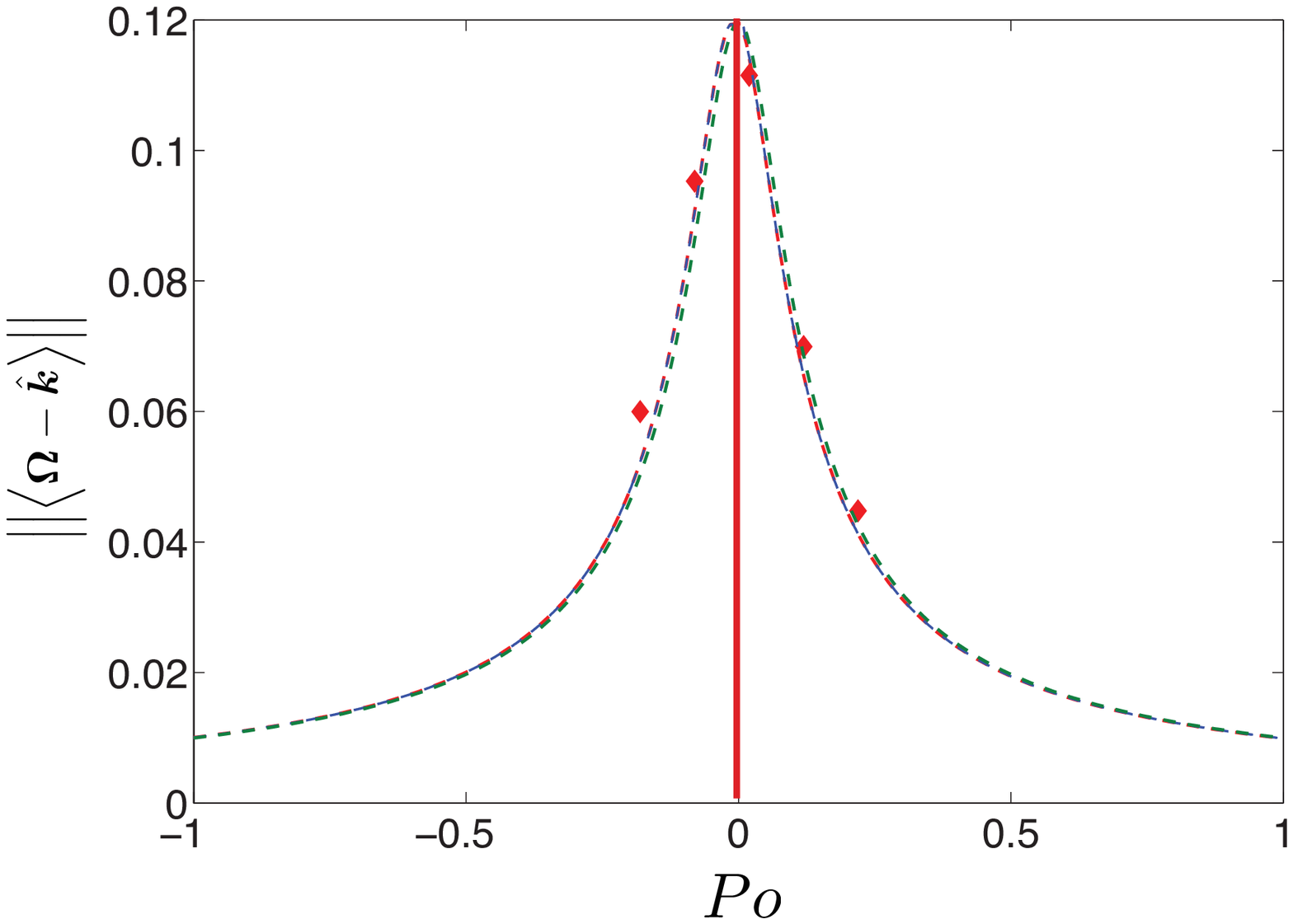}{Equatorial component of the fluid mean rotation as a function of the Poincar\'e number in a spherical geometry. The numerical simulation are performed with $E=10^{-3}$ and $Ro=10^{-2}$. The red symbols represents the simulations, the red dashed line represents the asymptotic solution of \citet{Busse1968}, the blue dashed line represents the solution of generalized model (\ref{eq_axi_P_x}-\ref{eq_axi_P_z}) and the dashed green line represents the reduced model (\ref{viscmeanvorteq1}-\ref{viscmeanvorteq3}) with a derived value of $\lambda=-2.62$. The red vertical line symbolized the region of the parameter space $\left| Po\right|<10^{-2}$ where no $\alpha$ can satisfy $Ro=Po\sin(\alpha)$ }{results_sphere}

In the present study we are concerned with the flow component of uniform vorticity. In the simulations, the uniform vorticity is obtained by averaging the fluid vorticity at each time step over a volume inside an ellipsoid:
\begin{equation}
\frac{x^2}{a^2}+\frac{y^2}{b^2}+\frac{z^2}{c^2}=d^2,
\end{equation}
with $d=1-5\sqrt{E}$ to exclude the Ekman boundary layer \cite[see also][]{cebron2010tilt}. 

\subsection{The axisymmetric spheroid, $a=b\neq c$}

In this particular geometry, the reduced model systematically leads to a flow steady in the precessing frame. 

\pict[10cm]{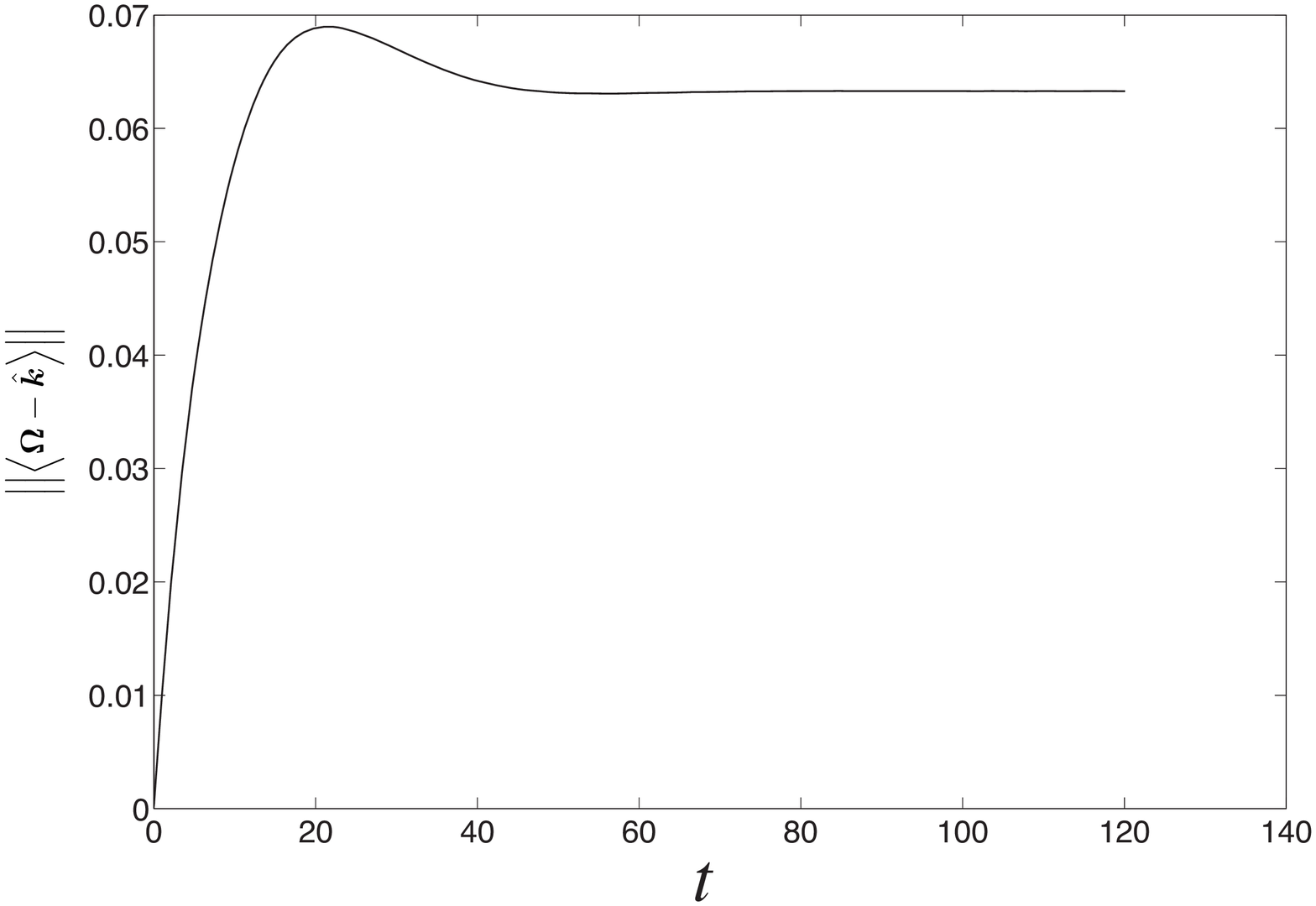}{Time evolution of the amplitude of the equatorial component of rotation of the fluid in the frame of precession from the simulations. $a=b=1$, $c=0.5$, $E=10^{-3}$, $Ro=10^{-2}$ and $Po=-0.45$.}{Omega_eq_ComSol_c05_Po045}

\pict[11cm]{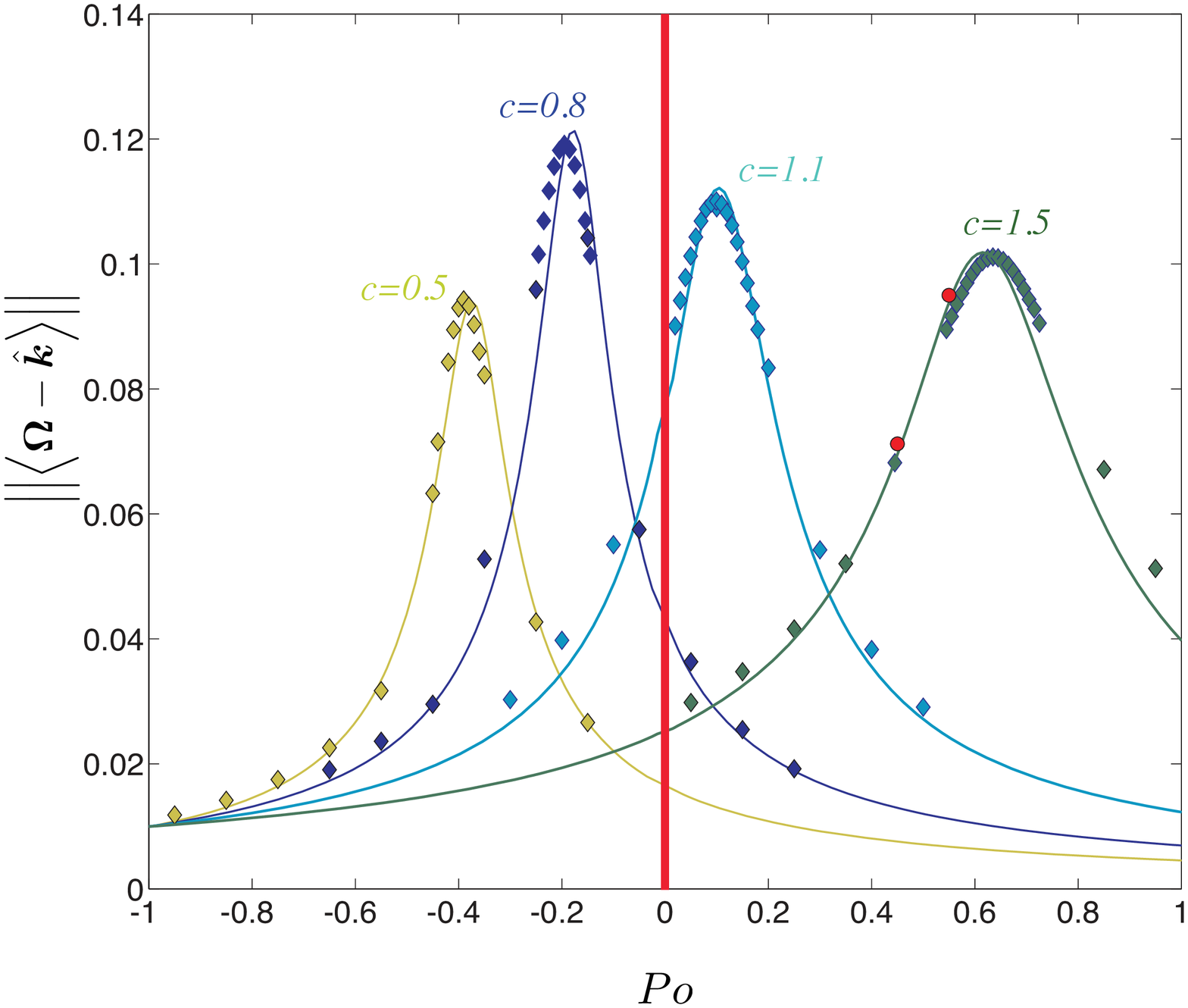}{Amplitude of differential rotation, $\normdw$, as a function of the Poincar\'e number. The symbols represent our numerical simulations simulations at $E=10^{-3}$ and $Ro=10^{-2}$, the solid lines represents the reduced model with the inverted values of $\lambda$ from table \ref{lambda_axi}. Each value of the polar axis c is represented by a different color as indicated. The red vertical line signifies the region of the parameter space $\left| Po\right|<10^{-2}$ where no $\alpha$ can satisfy $Ro=Po\sin(\alpha)$. The red circles represent simulations with c=1.5 with a spatial resolution four times larger ($135000$ DoF).}{Omega_Spheroid_Axi_comp_busse_generalizede}

Figure \ref{Omega_eq_ComSol_c05_Po045} represents the time evolution of the norm of the differential rotation, $\normdw$ from the three-dimensional (3D) non linear simulations ($a=b=1$, $c=0.5$, $E=10^{-3}$, $Ro=10^{-2}$ and $Po=-0.45$). It shows that the uniform vorticity component becomes stationary after a typical period of $60$ rotations, which is comparable to the spin-up time $t\propto E^{-1/2}\sim30$. This result is generic to all of our simulations in an axisymmetric spheroid. Hence, it validates the otherwise assumed stationarity of the uniform vorticity solution in the asymptotic theory of \citet{Busse1968} and \citet{Noir2003}.

Figure \ref{Omega_Spheroid_Axi_comp_busse_generalizede} shows the differential rotation, $\normdw$ for $a=b=1$ and $c=0.5/0.8/1.1/1.5$ as we scan in Poincar\'e numbers from $-1$ to $+1$. For each geometry, we perform a least squares inversion using the reduced model to determine the value of $\lambda$ that best fits the numerical simulations. The results for each value of $c$ are presented in table \ref{lambda_axi}. We choose to study each individual component in the precessing frame where the total vorticity remains time independent. Figure \ref{3Comp_axi} shows the individual components of $\vect{\Omega}$ viewed from the precessing frame.

We retrieve the classical result that the amplitude of the differential rotation, $\normdw$, exhibits resonant like peaks for a critical value of the Poincar\'e number, $Po_c$. Considering each individual component (Figure \ref{3Comp_axi}), the peaks correspond to a maximum of $\Omega_y$ and a (usually abrupt) change of sign of $\Omega_x$. As we shall see later at lower Ekman number, the term \textit{resonance} may have a significance in the inviscid limit but for finite viscosity we prefer to use the term \textit{transition} and define $Po_c$ as $\Omega_x(Po_c)=0$, the transition thus corresponding to the abrupt change of direction of the mean rotation axis of the fluid. Physically, $Po_c$ is the Poincar\'e number for which the equatorial component of the fluid rotation is exactly aligned with the gyroscopic forcing $\unit{k}_p\times\unit{k}$, leading to a pseudo-resonance between the precessional forcing and the so-called Poincar\'e mode (see \citet{Noir2003} for more details). As expected from the asymptotic and inviscid theory, $Po_c<0$ for an oblate spheroid, $a>c$, and $Po_c>0$ for a prolate spheroid, $a<c$. 

Figures \ref{3Comp_axi} shows the components of $\vect{\Omega}$ (viewed from the frame of precession). We compare the three different models, \citet{Busse1968}, the generalized model and the reduced model to the numerical simulations. Comparing the relative amplitude of the different components, the differential motion is clearly dominated by the equatorial component, which thus governs the evolution of $\normdw$ shown in figure \ref{Omega_Spheroid_Axi_comp_busse_generalizede}. We observe a quantitative agreement between the reduced model and the numerical results for $\Omega_x$ and $\Omega_y$. The small departure of $\Omega_z$ from $1$ is less accurately captured by the model, owing  to its weak influence on $\normdw$ from which we invert for the unique adjustable parameter $\lambda$. Meanwhile, the generalized model, without any adjustable parameter, predicts correctly the resonance positions but tends to overestimate the amplitudes as $c$ is increased. One can however notice that the results of this predictive model are still acceptable for $c \in [0.5 ; 1.1]$. In contrast, the usual \cite{Busse1968} model does not predict correctly the flow components' evolution, or even the resonance locations, as soon as the spheroid deformation becomes significant, which is expected given the domain of validity of this model.

\pict[9cm]{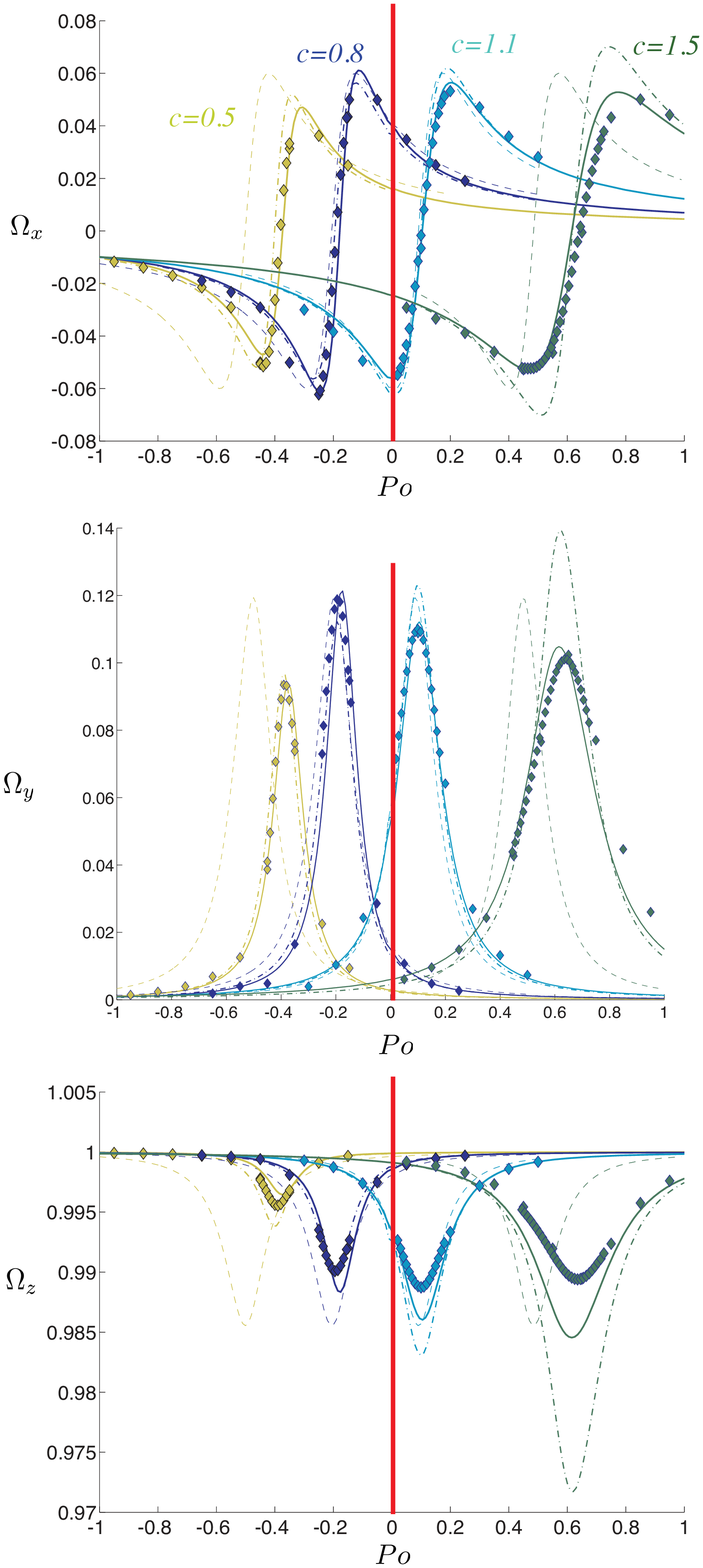}{From top to bottom: X,Y and Z components of the fluid rotation vector in the frame of precession within an axisymmetric spheroid. We compare our simulations (symbols), the theory of \citet{Busse1968}, represented by the dashed lines, our generalization of this model (dot-dashed lines), and the proposed reduced model (solid lines) with the inverted values of $\lambda$ from table \ref{lambda_axi}. Each value of the polar axis c is represented by a different color as indicated. The red vertical line symbolized the region of the parameter space $\left| Po\right|<10^{-2}$ where no $\alpha$ can satisfy $Ro=Po\sin(\alpha)$. }{3Comp_axi}

\begin{table}
\begin{center}
\caption{\label{lambda_axi} Inverted viscous coefficient $\lambda$ for an axisymmetric spheroid}
\begin{tabular}{|c|c|c|c|c|c|c|c|}
  c & 0.5 & 0.8 & 1.1 & 1.5 \\
  $\lambda$ & -3.35 & -2.57 & -2.78 & -2.97 \\ 
\end{tabular}
\end{center}
\end{table}

\subsubsection{The non-axisymmetric spheroid, $a \neq b=c=1$}

Figure \ref{time_evolution_non_axi} represents the time evolution of the three components of the fluid rotation vector in the frame of precession from the 3D non linear numerical simulations with $a=0.5,\,b=c=1$. Comparison with figure \ref{Omega_eq_ComSol_c05_Po045} shows clearly an important difference: in non-axisymmetric ellipsoids, an unsteady and periodic flow can be forced by the precession contrary to the flow forced in a spheroid which is steady. The inset shows moreover that $\Omega_x$ and $\Omega_y$ oscillate in phase quadrature, with the same amplitude $\delta/2$ and the same period, which is half the container rotation period $T_0$ (dimensionless value of $T_0$ is $2 \pi$).

Figure \ref{time_evolution_non_axi_3D} represents the same data set as in Figure \ref{time_evolution_non_axi} plotted in three dimensions to illustrate the dynamics of the mean rotation vector. The fluid rotation vector performs a time periodic quasi circular motion (red dots) around its mean position (blue arrow). The semi-aperture angle of the cone is given by $\sqrt{\Omega_x^2+\Omega_y^2}$.

\pict[13cm]{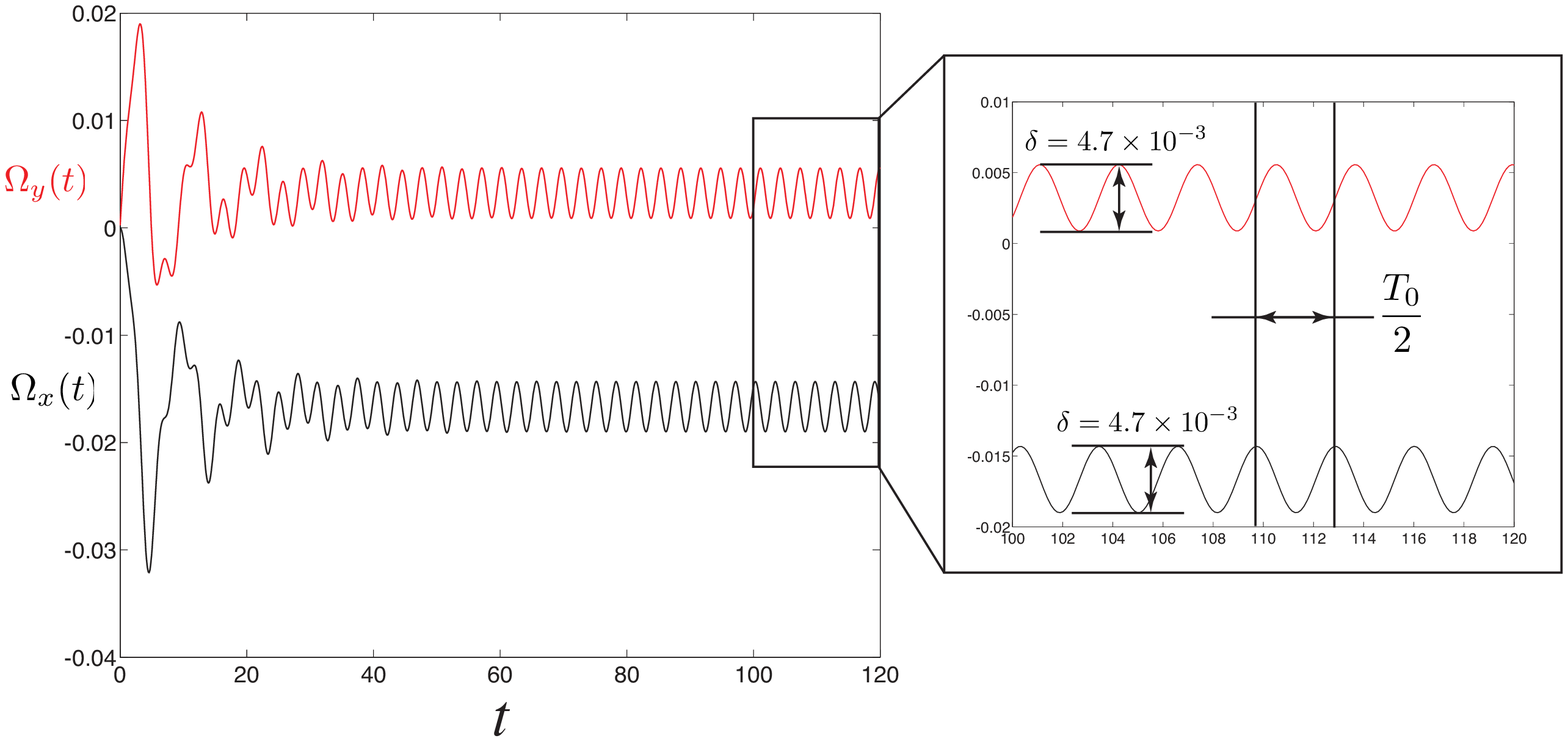}{Time evolution of the three component of the fluid rotation vector in the frame of precession $\Omega_x(t)$, $\Omega_y(t)$ from the numerical simulations for $a=0.5,\,b=c=1$, $E=10^{-3}$, $Ro=10^{-2}$ and $Po=-0.45$.}{time_evolution_non_axi}

\pict[10cm]{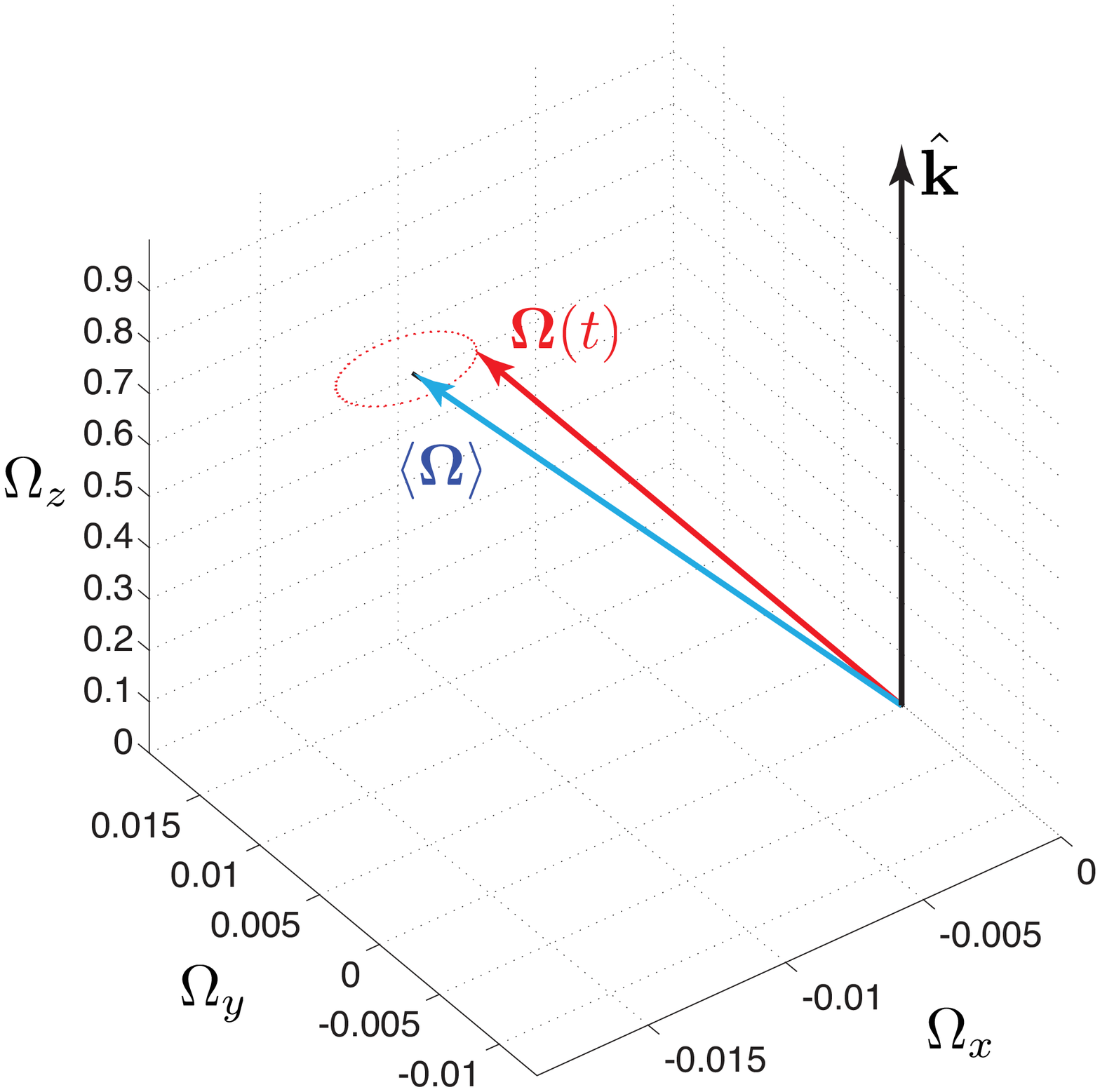}{Time evolution of the fluid rotation vector $\vect{\Omega}$ viewed from the frame of precession (Same set of parameters as in Figure \ref{time_evolution_non_axi}). The blue arrow represents the mean rotation vector $\left< \vect{\Omega} \right>$, the red arrow represents the instantaneous rotation vector at a particular time, the red dots show the trace of the instantaneous rotation vector for $60<t<120$ and the black arrow shows the container rotation vector.}{time_evolution_non_axi_3D}

We carry out a series of 3D numerical simulations for various geometries with $a \neq b=c$, in each case we perform the least squares inversion to determine $\lambda$ using only the time averaged differential rotation, $\normdwmean$ (Figure \ref{NonAxi_weq}). As in the case of an axisymmetric container, we observe peaks at critical values of the Poincar\'e number, identical in the mean and oscillatory components. We note that the critical $Po$ is retrograde for $a>1$ and prograde for $a<1$, which correlates with the results obtained in an axisymmetric spheroid. Indeed, in any meridional cross section of the non axisymmetric cavity, the trace of the boundary is an ellipse with a polar axis shorter than the mean equatorial axis for $a>1$, similarl to an oblate spheroid, and longer than the mean equatorial axis for $a<1$, similarl to a prolate spheroid. As the geometry tends toward the sphere $(a=1)$, the amplitude of the peak of the oscillatory component vanishes, while the peak of the time averaged component converges toward the solution for the sphere as illustrated in Figure \ref{NonAxi_weq}.

We observe a quantitative agreement between our reduced model and the numerical simulations for both the mean and oscillatory components for all cases with $a>b=c$. For $a<b=c$, the reduced model captures correctly the dynamics of the time average equatorial rotation but exhibit significant discrepancy in the axial components. 

\begin{table}
\begin{center}
\caption{\label{lambda_Naxi} Inverted viscous coefficient $\lambda$ for an non axisymmetric spheroid}
\begin{tabular}{|c|c|c|c|c|c|c|c|}
  a & 0.5 & 1.1 & 1.5 & 2\\
   $\lambda$ & $-4.5 \pm 0.02$ & $-2.54 \pm 0.02$ & $-2.29 \pm 0.02$ & $-2.03 \pm 0.02$\\
\end{tabular}
\end{center}
\end{table}

 \pict[8cm]{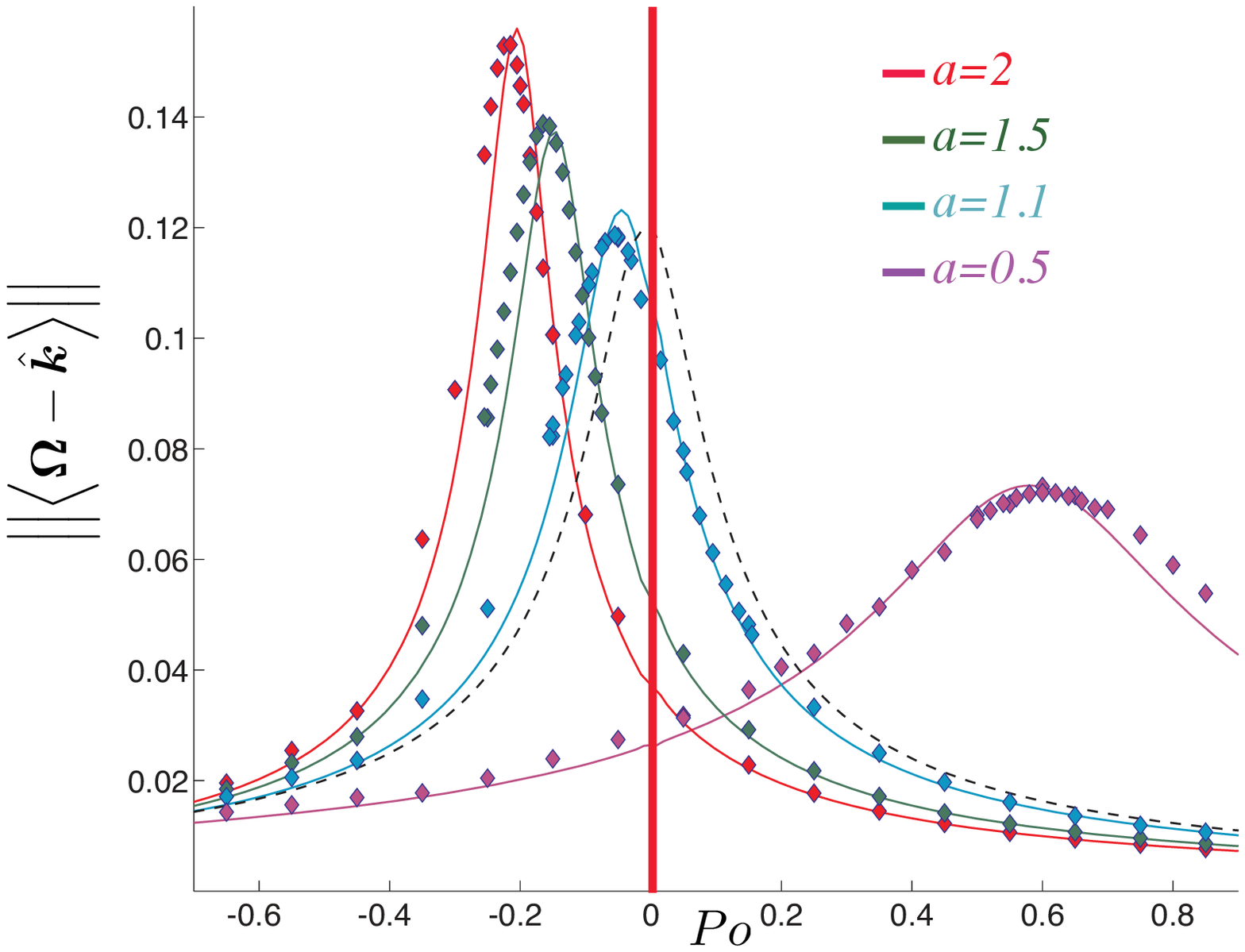}{Amplitude of the mean differential rotation, $\normdwmean$, as a function of the Poincar\'e number. The symbols represent numerical simulations at $E=10^{-3}$ and $Ro=10^{-2}$, the solid lines represent the inverted reduced model, the black-dashed line represents the solution for the sphere from \citet{Busse1968}. Each geometry, characterized by $a$, is represented with a different color as indicated. The red vertical line symbolized the region of the parameter space $\left| Po\right|<10^{-2}$ where no $\alpha$ can satisfy $Ro=Po\sin(\alpha)$.}{NonAxi_weq}
 
Figure \ref{3Comp_Nonaxi} shows the time averaged and oscillatory components, respectively, of $\vect{\Omega}$. The steady part of the uniform vorticity behaves as in an axisymmetric container, its axial component $\Omega_z$ departs only marginally from the vorticity of the container, hence, the differential motion between the fluid and the container is dominated by the equatorial component. Even though we invert for the unique adjustable parameter using the steady part only, we observe a very good agreement between the reduced model and the simulations. All three components exhibit a maximum amplitude at the critical $Po_c$ derived from the time average part. As suggested from the time evolution shown in Figure \ref{time_evolution_non_axi}, the two equatorial components have the same amplitude, the axial component is only five times smaller. In addition, we note a significant discrepancy both on the peak location and amplitude between the reduced model and the numerical simulations for $a=0.5$, similarly to the case of a prolate axisymmetric spheroid.

 \pict[13cm]{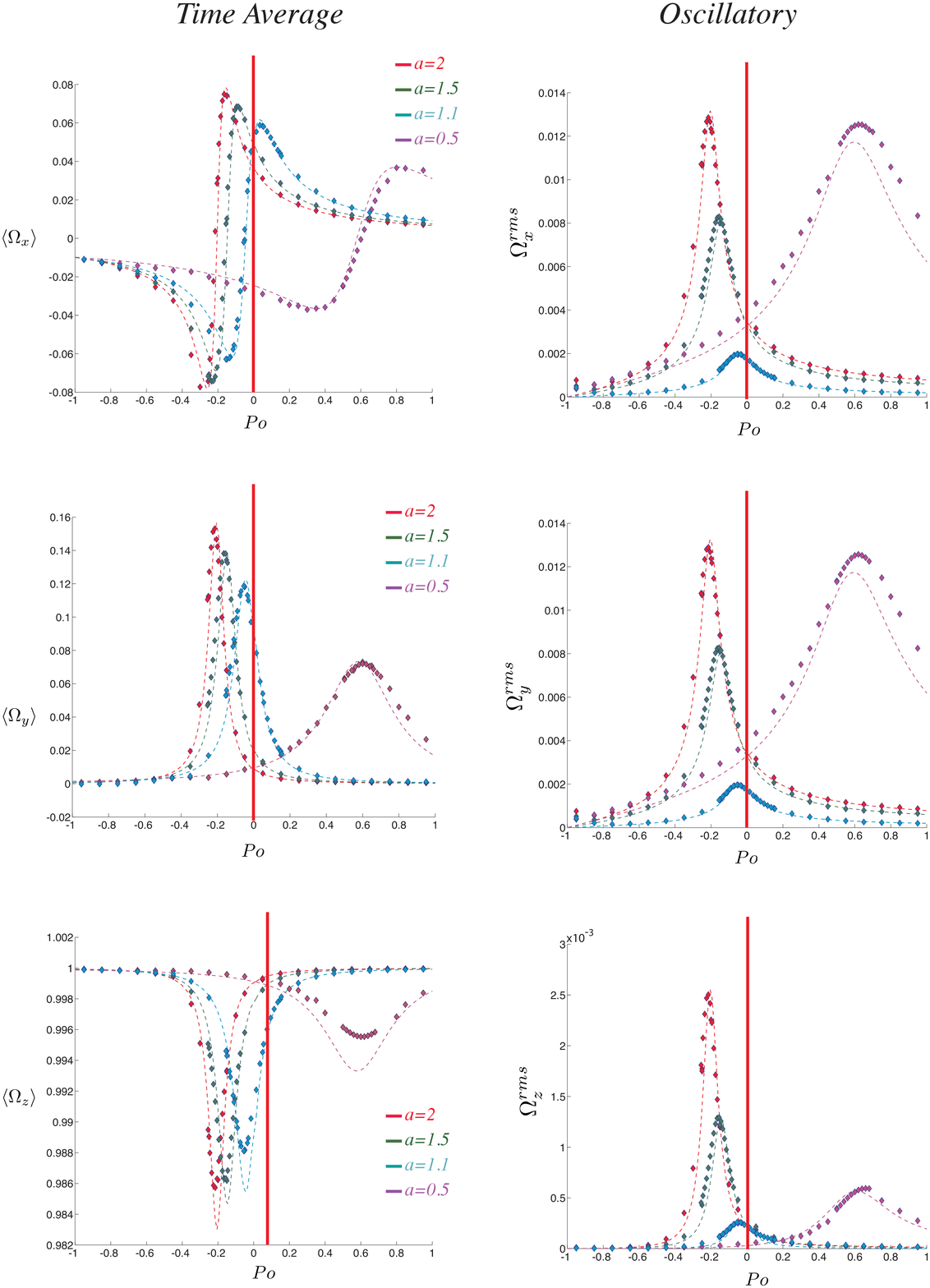}{From top to bottom: X,Y,Z-component of rotation of the fluid in the frame of precession. The left row shows the time averaged components, the right row shows the time standard deviation of the components. We compare, for each geometry (i.e. $a$), our simulations (symbols) and the reduced model (dashed lines) with the inverted values of $\lambda$ from table \ref{lambda_Naxi}. The red vertical line is the region of the parameter space $\left| Po\right|<10^{-2}$ where no $\alpha$ can satisfy $Ro=Po\sin(\alpha)$.}{3Comp_Nonaxi}

 \section{Discussion} \label{sec4}
In appendix \ref{compare_model_axi}, we show that the location of the $Po_c$, within an axisymmetric container, is determined primarily by the inviscid part of the equations, while the typical amplitude is most constrained by the decay rate $\lambda_{so}^r$. We also note that the viscous correction to the spin-over eigenfrequency accounts for a small shift of $Po_c$ but does not modify the fundamental dynamics, even at the moderate Ekman numbers considered here. Finally, the effect of the non vanishing axial differential rotation in the frame rotating with the fluid remains negligible in all of our simulation. The systematic mismatch of the amplitude of the generalized model with our simulations is likely due to the moderate Ekman numbers accessible in our numerical simulations. Meanwhile, the observed shift in $Po_c$ in Figure \ref{3Comp_axi} and Figure \ref{3Comp_Nonaxi} shows the limitations of the one adjustable parameter reduced model that only account for part of the dissipation mechanism.

The simulations presented here show that the flow of uniform vorticity in a non axisymmetric ellipsoid is not purely stationary in the frame of precession as it would be for a spheroidal cavity. This is supported by the governing equations (\ref{eq_Naxi_P_x}-\ref{eq_Naxi_P_z}) from which one can anticipate that, if a stationary uniform vorticity component exists, it will necessarily drive a time dependent perturbation for $\chi \neq 1$, i.e. when the two equatorial moments of inertia are not equal.

Our results suggest that the simple form of the viscous term (\ref{addhocvisc}) captures well the fundamental dynamics of the uniform vorticity flow in a non axisymmetric precessing ellipsoid. Taking advantage of the computational efficiency of this reduced model, we perform a series of time integrations at lower Ekman numbers.  Figure \ref{convergence_Ekman} shows the norm of the mean and oscillatory components of the differential $\vect{\Omega}-\unit{k}$ as a function of the Poincar\'e number for decreasing Ekman numbers in the case $a=1.5,\,b=c=1$; we assume the decay rate $\lambda$ to be independent of the Ekman number and equal to the value inverted in this geometry at $E=10^{-3}$ (see table \ref{lambda_Naxi}). As the Ekman number is reduced, both the stationary and the oscillatory part of the differential rotation tend toward an asymptotic limit already captured at $10^{-7}$.

\pict[18cm]{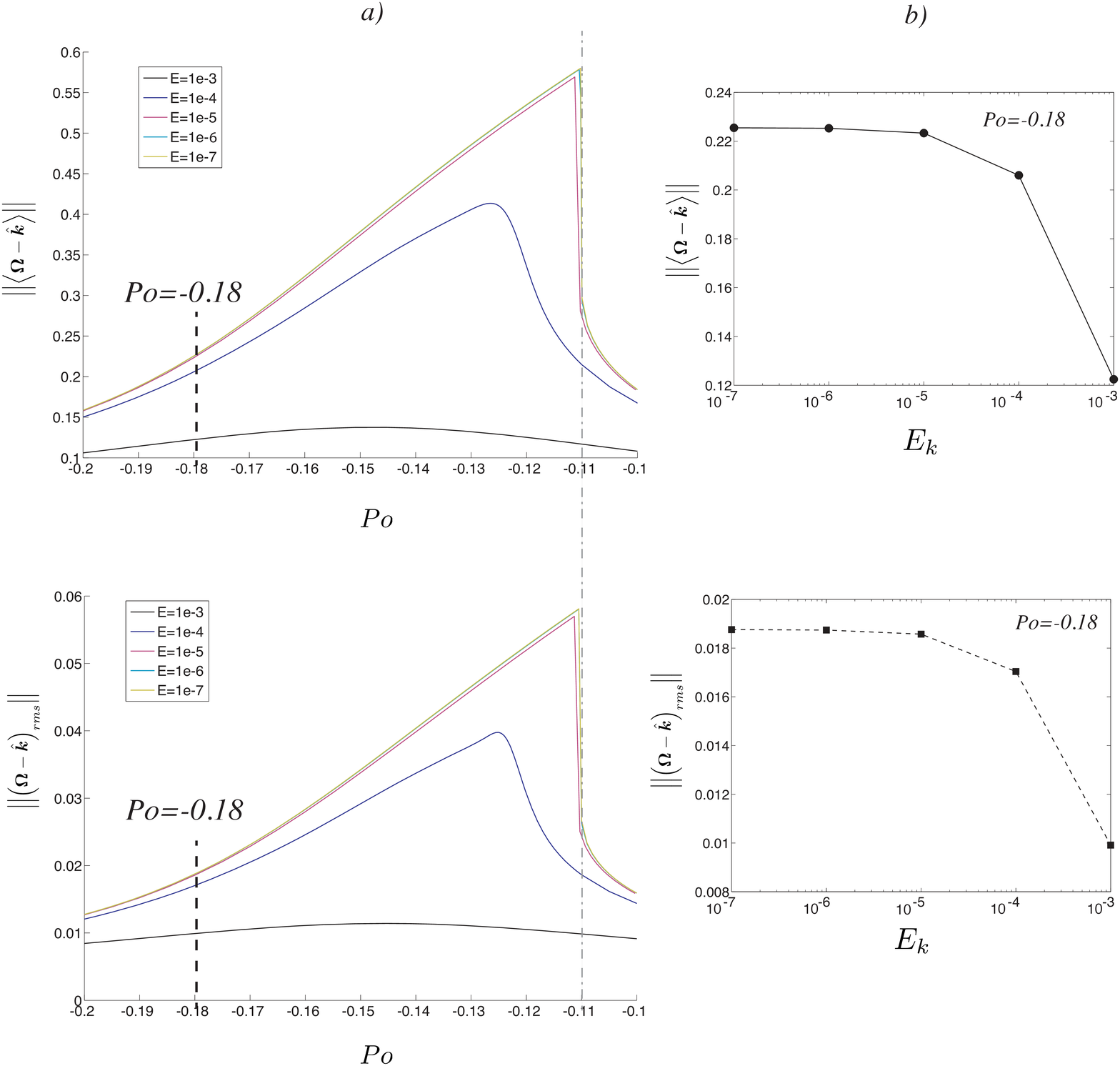}{a) Upper panel: norm of the stationary component of the differential rotation $\vect{\omega}$ as function of the Ekman number. Lower panel: norm of the oscillatory component of the differential rotation $\vect{\omega}$ as function of the Ekman number. b) Norm of the stationary (upper) and oscillatory (lower) part of the differential rotation for a fixed $Po=-0.18$ as a function of the Ekman number. In all calculations we integrate in time the reduced model with $a=1.5,\,b=c=1$ and $\lambda=-2.3$.}{convergence_Ekman}

Let us define the mean longitude $\phi$ and the mean latitude $\theta$ of the fluid rotation axis as follows:
\begin{eqnarray}
\cos\phi&=&\frac{\left< \Omega_x \right>}{\sqrt{\left< \Omega_x \right>^2+ \left<\Omega_y \right>^2}},\\
\tan\theta&=&\frac{\left< \Omega_z \right>}{\sqrt{\left< \Omega_x \right>^2+ \left<\Omega_y \right>^2}}.\\
\end{eqnarray}
Figure \ref{convergence_Ekman_direction} shows the evolution of the longitude and latitude for decreasing Ekman numbers. As for the amplitude, we observe that the direction of the stationary component of uniform vorticity tends toward an asymptotic value. We note that the asymptotic longitude is either $0^{\circ}$ or $180^{\circ}$, which corresponds to a fluid mean rotation vector lying in the plane $(\unit{k},\unit{k}_p)$. Hence, similarly to the case of an axisymmetric spheroid, it is the viscosity that forces the mean rotation vector to leave the plane containing the axis of the container and the axis of precession. In that plane, at vanishing Ekman numbers, the rotation vector evolves from high latitudes ($\vect{\Omega}$ almost aligned with $\unit{k}$), far from the transition, to mid latitudes near the transition. 

Our results suggest that at low enough Ekman number, the flow of uniform vorticity driven by the precession of the container becomes independent of $E$, or as a matter of fact, independent of the viscous term $\lambda \sqrt{E}$. Hence, outside of the transition region, the asymptotic solution for vanishing viscosity can be found using any arbitrary order $\mathcal{O}(1)$ value of the damping factor, providing that the Ekman number is small enough, typically when $E^{1/2}\lesssim(1-\chi)$.

\pict[14cm]{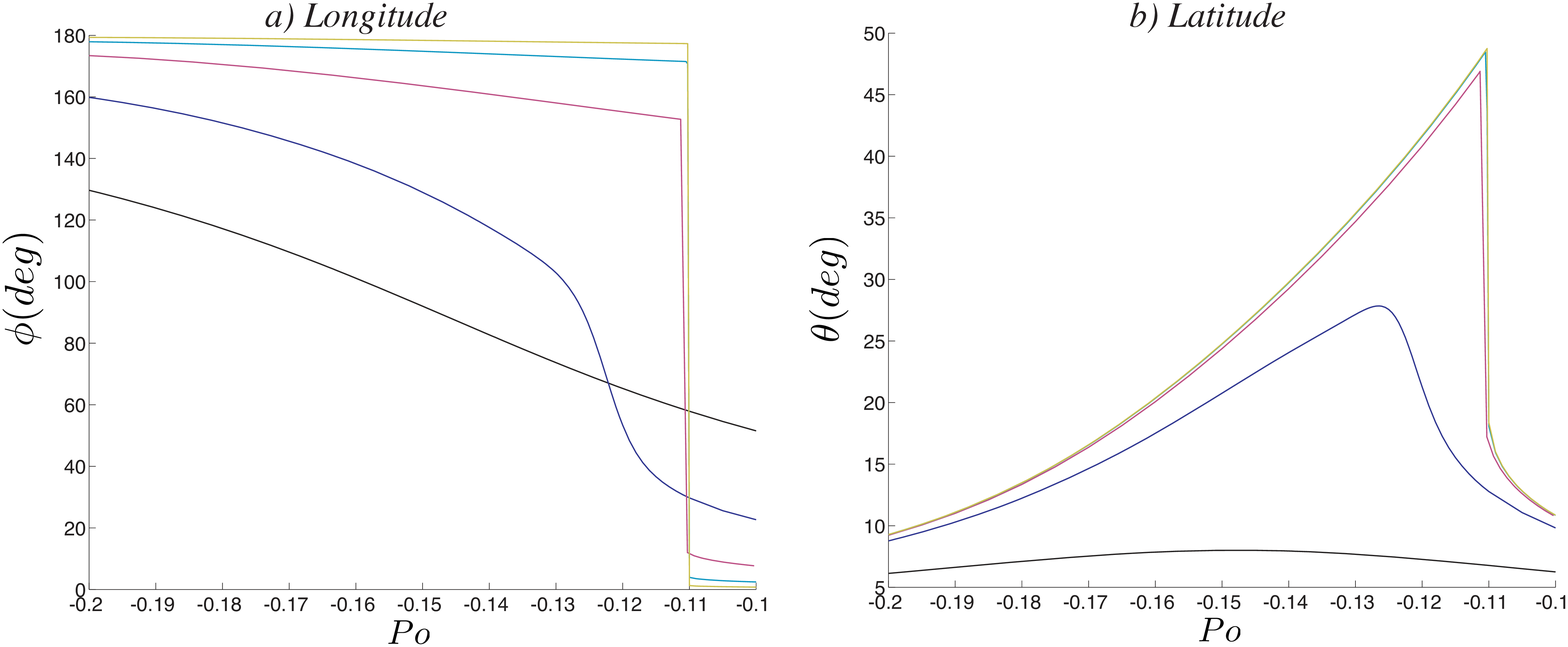}{a) Longitude of the stationary part of the rotation vector. b) Latitude of the stationary part of the rotation vector. Color code: same as Figure \ref{convergence_Ekman}.}{convergence_Ekman_direction}

\section{Conclusion} \label{sec5}
In the present study, we investigate the flow of uniform vorticity driven by precession in a spheroid and a non axisymmetric ellipsoid. We report the first numerical simulations in a non axisymmetric ellipsoid showing that, in contrast to a spheroid, the flow of uniform vorticity viewed from the frame of precession is no longer stationary.

In addition, we develop a semi-analytical model by first deriving the inviscid equations and then by reintroducing the viscosity. We propose a generalized model in the case of a spheroid of arbitrary ellipticity following the torque approach introduced by \citet{Noir2003} and using the linear asymptotic theory of the spin-over mode as a proxy. For non axisymmetric ellipsoids an analoguous theory has yet to be established and the same approach is not possible. Nevertheless, we introduce a reduced model with one adjustable parameter that we compare successfully with 3D non-linear numerical simulation at a fixed Ekman number, $E=10^{-3}$ (using the commercial software Comsol). 

Despite its simplicity, the reduced model with one adjustable parameter allows us to reproduce quantitatively the uniform vorticity flow obtained from numerical integrations of the full Navier-Stokes equations both in a spheroid and a non axisymmetric ellipsoid. Furthermore, the generalized model for a spheroid allow us to extend the classical asymptotic theory of \citet{Busse1968,Noir2003} to finite ellipticity as it is usually the case in laboratory experiments. With our current limited number of geometrical configurations (4 different values of a) it is not possible to ascertain the functional relationship between the geometrical deformation and the damping factor. 

Taking advantage of the computational efficiency of the reduced model compared to the full simulations of Navier-Stokes equations, we investigate the uniform vorticity flow in non axisymmetric ellipsoids as the Ekman number is decreased. We show that the uniform vorticity converges toward an asymptotic solution independent of the Ekman number and thus of the viscous term as a whole. At very low Ekman numbers, the time averaged component of the fluid rotation axis lies in the plane formed by the precession and container rotation vectors as in the case of a spheroid, meanwhile the time dependent component tends toward a finite amplitude.

\begin{table}
\begin{center}
\caption{\label{paramMoon} Dimensionless parameters for the Earth's moon.}
\begin{tabular}{|c|c|c|}
 \hline
$1-\chi$ & $2.5\times10^{-5}$ & \citet{lebarsNature}\\
$E$ & $\sim10^{-12}$ & \citet{lebarsNature}\\
$Po$ & $-3.9\times10^{-3} $ & \citet{meyer2011precession}\\
$\alpha$ & $1.54$ & \citet{meyer2011precession}\\
\hline
\end{tabular}
\end{center}
\end{table}

When looking at the dynamics in greater details we identify some limitations of our reduced model: first, there is a small shift of the critical value of the Poincar\'e number at which we observe a transition, and second the axial component of the fluid rotation exhibits a simpler dynamics in our reduced model than in the numerical simulations. 

Due to the limited range of Ekman numbers accessible in the numerical simulations, we believe that an experimental survey is necessary to complement the results presented in this study. With an experimental setup using water as a working fluid, a typical length scale $\sqrt{abc}\sim15$ cm  and rotating at $\Omega_0\sim240$ rpm, the achievable Ekman number will be of order $3\times10^{-6}$. Aside from testing the validity of the reduced model, it would be suitable to investigate the stability of the flow. 

Two recent publications have investigated possible mechanisms to drive Earth's Moon early dynamo. One invokes a precession driven turbulence in the liquid core \citep{dwyerNature}, whereas the other one proposes a meteoritic impact leading to a desynchronization of the Moon \citep{lebarsNature}. Considering our current tidally-locked Moon, the core-mantle boundary (CMB) geometry is close to a non-axisymmetric ellipsoid rather than a axisymmetric spheroid (typically, its shape can be approximated by the well known relation $(b-c)/(a-c)=1/4$ assuming hydrostatic state and homogeneous material). We can thus compare our reduced model, assumed to be valid for our current Moon, and the model of \citet{Busse1968}, which is valid at planetary settings but assumes a spheroidal CMB. The parameters used for the simulation are given in table \ref{paramMoon}. The reduced model for a non axisymmetric ellipsoid and the model of \citet{Busse1968} for a spheroid agree within $0.3\%$ leading to a mean differential rotation amplitude of order $3\%$ of the planet rotation rate and a core spin vector normal to the ecliptic plane in agreement with a former model by \citet{goldreich1967precession}. Neither the viscous nor the pressure torques are large enough to force the lunar core to precess with the mantle. Nevertheless, in contrast with a spheroidal model, the reduced model predicts an unsteady component of uniform vorticity of order $2.3\times10^{-6}\, \Omega_o$ oscillating with a period of $T\approx 13.5$ days. Although this amplitude is small, one may question what could happen if in the frame of precession there exists another source of gravitational perturbation at that frequency. Indeed, in that case, direct or parametric resonances may occur, leading to much larger amplitude flows, subsequent instabilities, and thus to enhanced dissipation.

\begin{acknowledgements}
The first author would like to dedicate this paper to the memory of Roland Leger (1921-2012). The authors would like to thank N. Rambaux for fruitful discussions on the initial work of H. Poincar\'e and J. Laskar for its invitation to the Poincar\'e 100 years anniversary which led to this study.
J. Noir is supported by ERC grant (247303 MFECE), D. C\'ebron is supported by the ETH Z\"urich Postdoctoral fellowship Progam as well as by the Marie Curie Actions for People COFUND Program. 
\end{acknowledgements}

\begin{appendix}

\section{The viscous torque for a precessing spheroid of arbitrary ellipticity} \label{appendixA}
\subsection{The viscous torque}\label{viscoustorquegeneral}
In the limit of small ellipticity, small Ekman number and small $Po\sin\alpha$, \citet{Busse1968} and \citet{Noir2003} have derived the viscous equations for the stationary flow of uniform vorticity in a precessing axisymmetric spheroid. We herein refer to this model as Busse 1968, who was the first one to derive it in the limit of small ellipticity, small Ekman number and small $Po\sin(\alpha)$. In this appendix, we follow the same approach as \citet{Noir2003} to derive a more general model for finite ellipticity. 
 
To reintroduce the viscosity, we assume a small Ekman number such that, at leading order, the uniform vorticity solution in the bulk remains essentially inviscid and the viscous forces are important only in the Ekman boundary layer. The Navier-Stokes equation for an arbitrary viscous flow $\vect{u}$ in the frame of precession leads to the following torque balance in the precessing frame (within the spheroid volume $V$):
\begin{eqnarray}
\overbrace{\int_V \boldsymbol{r} \times \frac{ \partial \vect{u}}{\partial t} \textrm{d} V}^{\boldsymbol{\Gamma_t}}+
\overbrace{\int_V \boldsymbol{r} \times (\vect{u}\cdot\nabla \vect{u}) \textrm{d} V}^{\boldsymbol{\Gamma_{nl}}}+
\overbrace{2 \int_V \boldsymbol{r} \times ( \boldsymbol{\Omega_p}
\times \vect{u}) \textrm{d} V}^{\boldsymbol{\Gamma_i}}= \nonumber\\
 \overbrace{-\int_V \boldsymbol{r} \times \nabla p \textrm{d}V}^{\boldsymbol{\Gamma_p}} + \overbrace{E \int_V
\boldsymbol{r} \times \nabla^2 \vect{u}
\textrm{d}V}^{\boldsymbol{\Gamma_v}}. \label{torque_t}
\end{eqnarray}

The challenge is thus to obtain the viscous torque due to the Ekman layer. 
 
As previously, we consider a uniform vorticity flow in a spheroid, which can be seen as a quasi solid body rotation along an axis tilted from the container rotation axis. Note that no further assumption is made on the stationarity in the frame of precession. For the particular flow $\vect{U}=\vect{\omega}\times \vect{r}+\vect{\nabla}\phi$, the integration of $\vect{\Gamma_t}$, which is carried in the coordinates system attached to the ellipsoidal container, leads to:

\begin{equation}
\mathcal{L} \vect{\Gamma}_t=\frac{\partial \vect{\omega}}{\partial t},
\end{equation}
where $\mathcal{L} $ is the matrix:

\begin{eqnarray}\label{transform}
\mathcal{L} =
 \frac{15}{16\pi}  \left [
   \begin{array}{ccc}
      \frac{b^2+c^2}{b^2c^2} & 0& 0 \\
      0& \frac{a^2+c^2}{a^2c^2} & 0 \\
      0& 0& \frac{b^2+a^2}{b^2a^2} \\
   \end{array}
   \right ].
\end{eqnarray}

The differential rotation between the fluid and the surrounding container in the frame can be decomposed into an axial and an equatorial component relative to the rotation axis of the fluid:
\begin{eqnarray}
\vect{\delta \omega}_{z}&=&\left( \frac{\vect{\Omega}-\unit{k}}{\Omega^2}\cdot \vect{\Omega}\right)\vect{\Omega},\\
\vect{\delta \omega}_{eq}&=& \vect{\Omega}-\unit{k} -\vect{\delta \omega}_{z}.
\end{eqnarray}
Without the viscous torque acting on the fluid, the equatorial component would tend to grow a spin-over mode, while the axial component would result in a spin-up or spin-down of the fluid. Thus, following the approach of \citet{Noir2003}, the viscous torque can be derived from the linear decay rate of Greenspan for the spin-over and spin-up: 
\begin{eqnarray}
\mathcal{L} \vect{\Gamma}_{\nu}^{eq,z}=\left. \frac{\partial (\vect{\delta \omega}_{eq,z})}{\partial t}\right|_{t=0}.
\end{eqnarray}
Since the linear calculation is only valid in the frame rotating with the fluid, we introduce a modified Ekman number $E_f=E/\Omega$ and a rescaled time $t_f=\Omega t$ associated to this frame of reference. According to \cite{Greenspan1968}, the time evolution of the spin-over mode in the non-rotating frame can be written as \citep{Noir2003}:
\begin{eqnarray}
\vect{\delta \omega}_{eq}(t)=\exp \left( \lambda^r_{so}E_f^{1/2}t_f\right) & & \left[ \vect{\delta \omega}_{eq}(0)\, \cos \left( \lambda^i_{so}E_f^{1/2}t_f\right) \right. \nonumber\\
& & - \left. \vect{\Omega}\times\vect{\delta \omega}_{eq}(0)\, \sin \left( \lambda^i_{so}E_f^{1/2}t_f\right) / \Omega \right].
\end{eqnarray}
It follows:
\begin{eqnarray}\label{viscousSO}
\mathcal{L} \vect{\Gamma}_{\nu}^{eq}=(E\Omega)^{1/2}\left[ 
\frac{\lambda^r_{so}}{\Omega^2}\left( 
 \begin{array}{ccc}
      \Omega_x\Omega_z \\
      \Omega_y\Omega_z\\
      \Omega_z^2-\Omega^2\\
   \end{array}
   \right)+
 \frac{\lambda^i_{so}}{\Omega}\left(   
  \begin{array}{ccc}
     \Omega_y\\
     -\Omega_x\\
      0\\
   \end{array}
   \right)
   \right].
\end{eqnarray}
In contrast with \citet{Noir2003}, who use the $\lambda^r_{so},\,\,\lambda^i_{so}$ in the spherical approximation of \citet{Greenspan1968}, we propose to use the analytical prediction of $\lambda^r_{so},\,\,\lambda^i_{so}$ obtained from \citet{Zhang2004} in an oblate spheroid ($c<a$) of arbitrary ellipticity. Although the author does not claim that his derivation remains valid in a prolate spheroid ($c>a$), we have checked that the formula reproduces the results of \cite{Greenspan1968} and are thus valid for prolate spheroids. It is important to note that the derivation of the spin-over decay rate are valid only for an axisymmetric container. Then, if the tilt of the fluid rotation axis and the ellipticity is not small enough, the viscous torque in the precessing cavity can no longer be inferred from the axisymmetric spin-over mode asymptotic theory introduced above. 

The axial differential rotation can be treated similarly. Without the viscous torque the axial differential rotation would tend to spin-up or spin-down the fluid. From \citet{Greenspan1968}, the time evolution of an axial differential rotation can be written as
\begin{eqnarray}
\vect{\delta \omega}_z(s,t)=\vect{\delta \omega}_z(0)\left( 1- \exp(\lambda_{sup}^*(s)\, E_f t_f)\right),
\end{eqnarray}
with a coefficient $\lambda_{sup}^*(s)$ which changes with the cylindrical radius $s$. An explicit analytical expression of $\lambda_{sup}^*(s)$ is given by \cite{greenspan1963time} for axisymmetric containers, which can be written in the case of a spheroid as:
\begin{equation}
\lambda_{sup}^*(s)= - \frac{[1-s^2 (1-c^2)]^{1/4}}{c\, (1-r^2)^{3/4}}.
\end{equation}
Hence, the axial viscous torque can be estimated as:
\begin{eqnarray}\label{viscousSUP}
\mathcal{L} \vect{\Gamma}_{\nu}^{z}=\lambda \sqrt{E}_{sup}\left(1-\frac{\Omega_z}{\Omega^2}\right)\left( 
 \begin{array}{ccc}
      \Omega_x\\
     \Omega_y\\
    \Omega_z\\
   \end{array}
   \right),
\end{eqnarray}
with
\begin{eqnarray}
\lambda_{sup}=\int \lambda_{sup}^*(s)\, \textrm{d}s = - \frac{\sqrt{\pi^3/2}}{c\, \mathrm{\Gamma}(3/4)^2}\, \mathrm{F}\left([-1/4, 1/2],[3/4], 1-c^2 \right),
\end{eqnarray}
where $\mathrm{\Gamma}$ is simply the gamma function and $\mathrm{F}(n,d,z)$ is the usual generalized hypergeometric function, also known as the Barnes extended hypergeometric function \cite[see respectively chap. 6 and 15 of][]{Abramovitz}. Note that when the tilt of the fluid mean rotation axis with the one of the mantle becomes large, i.e. when the container is no longer axisymmetric from the fluid point of view, we do not expect this derivation of the torque to apply either.  

Finally, taking into account both the spin-up and spin-over contributions, it yields
\begin{eqnarray}
\mathcal{L} \vect{\Gamma}_{\nu}= \sqrt{E\Omega} \left[ 
\frac{\lambda^r_{so}}{\Omega^2}\left( 
 \begin{array}{c}
      \Omega_x\Omega_z \\
      \Omega_y\Omega_z\\
      \Omega_z^2-\Omega^2\\
   \end{array}
   \right)+
 \frac{\lambda^i_{so}}{\Omega}\left(   
  \begin{array}{c}
     \Omega_y\\
     -\Omega_x\\
      0\\
   \end{array}
   \right)
   +
   \lambda_{sup} \frac{\Omega^2-\Omega_z}{\Omega^2} \left( 
 \begin{array}{c}
      \Omega_x\\
     \Omega_y\\
    \Omega_z\\
   \end{array}
   \right)
\right].
\end{eqnarray}

Substituting (\ref{change_xyz}) into the equations (\ref{viscmeanvorteq1}-\ref{viscmeanvorteq3}) with $a=b$, we obtain the viscous equations in the frame of precession:
\begin{eqnarray}
\frac{\partial \Omega_x}{\partial t}&=&P_z\Omega_{y}-(1-\gamma)\left[P_z\Omega_{y}+\Omega_{y}\Omega_{z}\right]+ \mathcal{L} \vect{\Gamma}_{\nu}\cdot \unit{e}_x, \label{eq_axi_P_x}\\
%
\frac{\partial \Omega_y}{\partial t}&=&P_x\Omega_{z}-P_z\Omega_{x}+(1-\gamma)\left[P_z\Omega_{x}+\Omega_{x}\Omega_{z}\right]+\mathcal{L} \vect{\Gamma}_{\nu}\cdot \unit{e}_y, \label{eq_axi_P_y}\\
%
\frac{\partial \Omega_z}{\partial t}&=&-P_x\Omega_{y}-(1-\gamma) P_x\Omega_{y}+ \mathcal{L} \vect{\Gamma}_{\nu}\cdot \unit{e}_z, \label{eq_axi_P_z}
 \end{eqnarray}
where $\gamma=(2a^2)/(a^2+c^2)$ represents the ration of the polar to equatorial moment of inertia. 

Taking (\ref{eq_axi_P_x})$\times \Omega_x+$ (\ref{eq_axi_P_y})$\times \Omega_y+$ (\ref{eq_axi_P_z})$\times \Omega_z$ yields,
\begin{eqnarray}
(\vect{\Omega}-\unit{k})\cdot\vect{\Omega}=\frac{(1-\gamma) P_x \Omega_y\Omega_z}{\lambda_{sup} \sqrt{E}}.
\end{eqnarray}
Then, in the limit $(1-\gamma)P_x / \sqrt{E} \ll 1$, we recover the so called no spin-up condition introduced by \citet{Noir2003}, also equivalent to the solvability condition of \citet{Busse1968}. This condition, also used by \cite{cebron2010tilt}, is thus not valid in general for a spheroid of arbitrary ellipticity.

\subsection{Comparison between the different models in an axisymmetric spheroid.}\label{compare_model_axi}
Substituting (\ref{change_xyz}) into the equations (\ref{viscmeanvorteq1}-\ref{viscmeanvorteq3}) with $a=b$, we obtain the viscous equations in the frame of precession. We thus have three different models for the axisymmetric spheroid: the asymptotic analysis of \citet{Busse1968}, our generalized model and our reduced model. The fundamental differences between all three models are twofold. First the model of \citet{Busse1968,Noir2003} uses an approximate form of the inviscid part of the equations, valid only for small departure from the sphere ($1-\gamma \ll 1$) and for small $Po \ll 1$, while the generalized and reduced models uses an exact derivation for the inviscid part. Second all three models are based on a different derivation of the viscous torque: \citet{Busse1968,Noir2003} are based on the asymptotic values of $\lambda_{so}^{i,r}$ of the sphere and on the no spin-up condition, the generalized model uses the asymptotic values of $\lambda_{so}^{i,r}$ for an oblate spheroid of arbitrary ellipticity from \citet{Zhang2004} and does not impose the no spin-up condition, and finally the reduced model neglects the terms proportional to $\lambda_{so}^i$ and $\lambda_{sup}$ and we thus have to close by fitting the best value of $\lambda_{so}^r$. 
 
Figure \ref{relative_cont_LrLiLsup_spheroid}(a) shows the contribution of the different terms of the viscous torque in the generalized model (\ref{viscousSO}-\ref{viscousSUP}). We observe that throughout the entire range of $Po$ and for all geometries, the terms proportional to $\lambda_{sup}$ (due to the axial differential rotation) remains two to three orders of magnitude smaller than the term proportional to $\lambda^r_{so}$ and can therefore be neglected. The contribution from the term proportional to $\lambda^i_{so}$ remains four to twenty times smaller than the term proportional to $\lambda^r_{so}$. Although not negligible, this term is expected to have a limited effect on the dynamics of the uniform vorticity flow. In most of our simulations, the generalized model reduces thus to the reduced model. This is illustrated in Figure \ref{relative_cont_LrLiLsup_spheroid}(b) which compares, for the axisymmetric spheroids considered in this study, the generalized and reduced models with the same value of $\lambda_{so}^r$. We observe a small shift in the peaks location which reflects the absence of the correction in $\lambda^i_{so}$ in the reduced model. In agreement with Figure \ref{relative_cont_LrLiLsup_spheroid}(a), this shift is larger for $c=0.5$, where both the $\lambda_{so}^r$ and $\lambda_{so}^i$ contributions are of the same order. 

\pict[15cm]{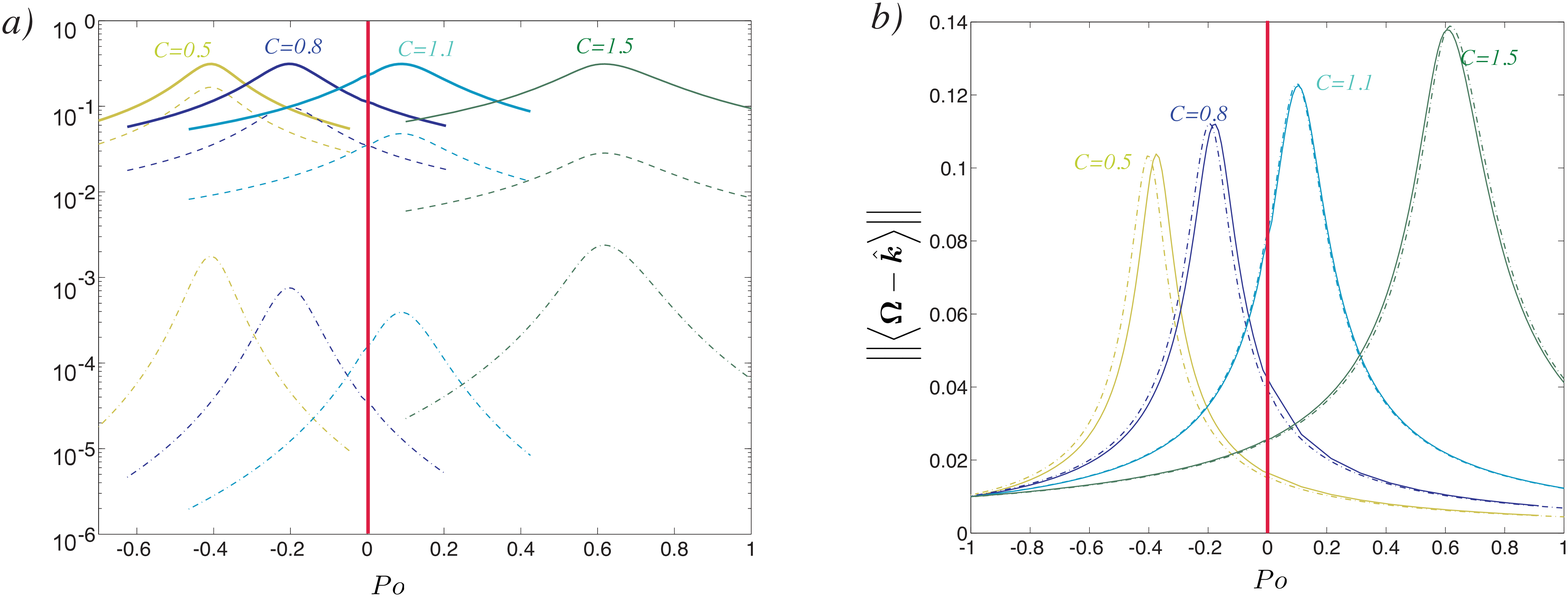}{a) Amplitude of the viscous terms in the generalized model associated with $\lambda^r,\,\lambda^i$ and $\lambda_{sup}$ (\ref{viscousSO}-\ref{viscousSUP}). The color scheme stands for the different polar flattening as indicated, the solid line represents the contribution from the $\lambda^r$-term, the dashed line represents the contribution from the $\lambda^i_{so}$-term and the dot-dashed line represents the contribution from the $\lambda_{sup}$-term. b) Comparison of the equatorial component of rotation between the generalized model (dot-dashed lines) and the reduced model (solid lines). In both models we use, $a=b=1$, $c=0.5/0.8/1.1/1.5$, $E=10^{-3}$, $Ro=10^{-2}$ and the values of $\lambda^r,\,\lambda^i$ of \citet{Zhang2004}. The red vertical line symbolized the region of the parameter space $\left| Po\right|<10^{-2}$ where no $\alpha$ can satisfy $Ro=Po\sin(\alpha)$.}{relative_cont_LrLiLsup_spheroid}

We now compare the generalized model and reduced models with $\lambda=\lambda_{so}^r=-3.03$ \citep{Zhang2004} to the asymptotic solution of \citet{Busse1968,Noir2003} using both the asymptotic value $\lambda=\lambda_{so}^r=-2.62$ \citep{Greenspan1968} and the asymptotic value $\lambda=\lambda_{so}^r=-3.03$ \citep{Zhang2004} (Figure \ref{effect_eq_lr_li}). In addition, we represent the $Po_c$ for the classical inviscid of Poincar\'e obtained by substituting $\mathcal{L}\boldsymbol{\Gamma_{\nu}}=0$ in (\ref{eq_axi_P_x}-\ref{eq_axi_P_z}) and assuming a stationary solution. It illustrates that the location of the peak is determined primarily by the inviscid form of the equations, that are exact in our model and approximated for small $Po$ and small ellipticity in \citet{Busse1968,Noir2003}. Meanwhile, as seen from our reduced model, the variation in $\lambda^i_{so}$ contributes to a small detuning of the peak but the amplitude is mostly determined by the decay rate $\lambda^r_{so}$.

\pict[10cm]{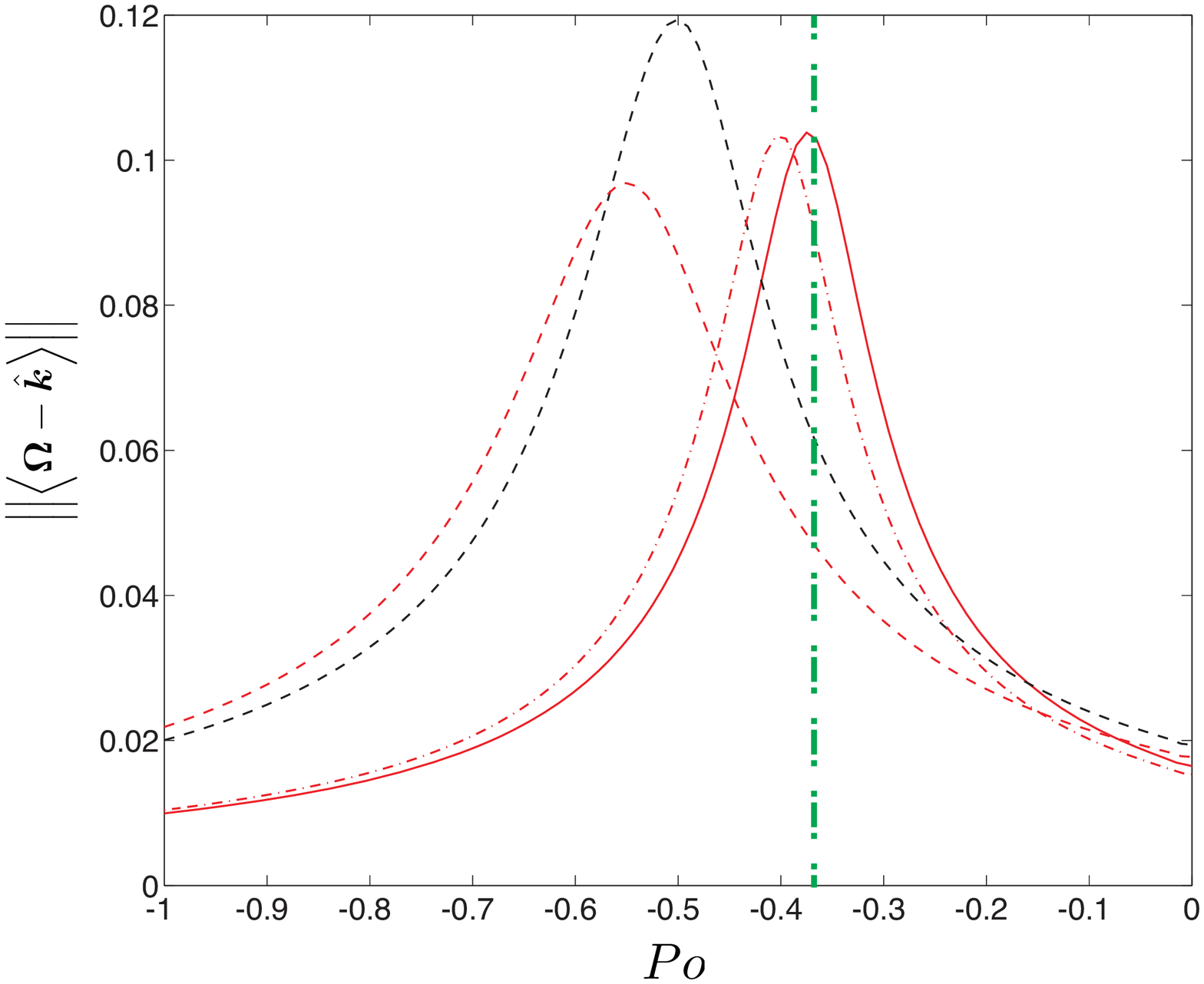}{Norm of the differential rotation for $a=b=1$, $c=0.5$, $E=10^{-3}$, $Ro=10^{-2}$ The solid lines represents the reduced models, the dashed line represents the asymptotic theory of \citet{Busse1968,Noir2003} and the dot-dashed line represents the generalized model. The color scheme stands for the different values of $\lambda_{so}^r,i$, from \citet{Greenspan1968} (black) and from \citet{Zhang2004} (red). The green dot-dashed line represents the critical Po predicted from a purely inviscid model.}{effect_eq_lr_li}

This validates the use of the reduced model in the case of an axisymmetric spheroid and we are confident that the same general remarks apply to the case of a non axisymmetric container.

\end{appendix}

\bibliographystyle{jfm}
\bibliography{reference_precession}

\end{document}